\documentclass[11pt,a4paper]{article}

\usepackage[english]{babel} 
\usepackage{dsfont} 
\usepackage{amsmath,amsfonts,amsthm}
\usepackage{a4wide}
\usepackage{fancyhdr}
\usepackage{latexsym}
\usepackage{amsmath,amssymb,amsfonts}%
\usepackage{paralist}
\usepackage{lscape}
\usepackage{url}

\usepackage{graphicx}%

\usepackage{amsthm}%
\usepackage{mathrsfs}%
\usepackage{bm}
\usepackage[title]{appendix}%
\usepackage{xcolor}%
\usepackage{textcomp}%
\usepackage{manyfoot}%
\usepackage{booktabs}%
\usepackage{algorithm}%
\usepackage{algorithmicx}%
\usepackage{algpseudocode}%

\usepackage[numbers,sort&compress]{natbib}
\usepackage{threeparttable}

\usepackage{float}
\usepackage{verbatim}
\usepackage{longtable}
\usepackage{tabularx}       
\usepackage{multicol}
\usepackage{caption}
\usepackage{floatflt}
\usepackage{afterpage}
\usepackage{moreverb}     
\usepackage{bbm}
\usepackage{booktabs}
\usepackage[colorlinks = true, linkcolor = blue, citecolor = blue, urlcolor = blue, breaklinks = true]{hyperref}

\usepackage{float} 

\usepackage{multirow}
\usepackage[margin=1in]{geometry}
\usepackage{calc}

\allowdisplaybreaks 

\AtBeginDocument{}   

\newcounter{saveeqn}

 \theoremstyle{definition}

\newtheorem{theorem}{Theorem}
\newtheorem{lemma}[theorem]{Lemma} 
\newtheorem{corollary}[theorem]{Corollary}
\newtheorem{proposition}[theorem]{Proposition}
\newtheorem{definition}{Definition}
\newtheorem{remark}{Remark}
\newtheorem{example}{Example}

\newcommand{\Var}{\text{Var}}

\usepackage{enumitem}

\begin{document}

\begin{center}
{\Large {\bf Expectiles as basis risk-optimal payment schemes \\in parametric insurance}}\\
\vspace*{1cm}
{\large Markus Johannes Maier\textsuperscript{1*} and Matthias Scherer\textsuperscript{1}}\\
\vspace*{0.5cm}
\textsuperscript{1}Chair of Mathematical Finance, Technical University of Munich (TUM), Parkring 11, D-85748 Garching-Hochbrück, Germany\\
\vspace*{0.5cm}
\textsuperscript{*}Corresponding author, e-mail: \href{mailto:markus.j.maier@tum.de}{markus.j.maier@tum.de}\\
Contributing author: \href{mailto:scherer@tum.de}{scherer@tum.de} 
\end{center}
\vspace*{0.4cm}

\begin{abstract}\noindent
Payments in parametric insurance solutions are linked to an index and thus decoupled from policyholders' true losses. While this principle has appealing operational benefits compared to traditional indemnity coverage, i.e. is very efficient and cost effective, a downside is the discrepancy between payouts and actual damage, called basis risk. We show that in an asymmetrically weighted mean square error framework, the basis risk-minimizing payment schemes for pure parametric and parametric index insurance contracts can be expressed as conditional expectiles of policyholders' true loss given a compensation-triggering incident. We provide connections to stochastic orderings and demonstrate that regression approaches allow easy implementation in practice. Our results are visualized in parametric coverage for cyber risks and agricultural insurance. 

\vspace{0.4cm}
\end{abstract}
 

\section{Introduction}\label{sec1}

Parametric insurance differs from traditional indemnity insurance primarily through the structure and timing of its payouts to policyholders. Technical details (deductibles, limits, co-insurance) notwithstanding, indemnity-based contracts compensate policyholders for experienced losses, the exact amount is determined after the respective event. In contrast, parametric insurance relies on a predefined, typically publicly observable parameter (vector) or index to govern payouts \cite{Louaas2024}. Policyholders receive payments once a trigger requirement, determined at contract inception, is met\footnote{Regulators might additionally require demonstration of insurable interest at the time of the incident, e.g., through policyholders verifying the occurrence of actual losses \cite{Lin2020}.}. While decoupling payments from policyholder-specific events yields certain benefits, e.g.\ swift and reliable compensation \cite{Jerry2023, Dirks2020}, it leaves customers exposed to basis risk. Although there exist various definitions for basis risk, see \cite{Clement2018}, they all refer to a mismatch between policyholders’ compensation $Y$ and their true loss\footnote{Approaches in agriculture index insurance may instead compare true and predicted yield for basis risk assessment (see, e.g., \cite{Boyd2020}). Interpreting this as comparing $S$ and $Y$ under a normalized crop market price of $1$, this still fits the above setting.} $S$. This mismatch is often assessed via the difference between $S$ and $Y$ (see, e.g., \cite{Chantarat2013, Dercon2014}) or in terms of the correlation between the underlying index $\theta$ and $S$ (see, e.g., \cite{Clement2018, Doherty2002}). Focusing on payout-based assessments, Table~\ref{BasisRiskMeasures} surveys basis risk metrics in existing research. 
\par In this work, we define basis risk as the squared difference between losses and compensation, where positive and negative deviations might be weighted differently. On the applied side, this takes into account important characteristics of basis risk. From the mathematical perspective, this generalizes the popular MSE metric and allows to be interpreted as a special case of basic penalty functions. Finally, we can embed basis risk minimization into the domain of expectiles. These are generalized quantiles, first defined by Newey and Powell \cite{Newey1987} in the context of asymmetric least squares regression. They have received much attention in the domain of risk measurement (see, e.g., \cite{Bellini2016} and the references therein), particularly for being the only elicitable coherent risk measure \cite{Bellini2015, Delbaen2016, Ziegel2016}. As a result, expectiles have been considered an easy-to-backtest (tail) risk measure (see, e.g., \cite{Girard2020, Daouia2018}) in the context of indemnity insurance. Further actuarial applications include premium calculation (as a representative of the zero utility premium principle \cite{Bellini2018} and through the ``expectile premium principle’’ \cite{Gao2023}) and optimal contract design in the presence of exclusions (see, e.g., \cite{Chi2017} and the references therein). To the best of our knowledge, however, our results establish the first connection between expectiles (and the generalized notion of conditional expectiles, see \cite{Bellini2018}) and the area of parametric insurance.

\begin{table}[t]
\begin{threeparttable}
\caption{List of basis risk metrics, comparing a loss $S$ to the parametric insurance payout $Y$, as proposed in existing literature. No differentiation is made between stochastic expressions and their empirical counterparts (e.g., between the $L^2$-distance and the mean squared error).}\label{BasisRiskMeasures}
\begin{tabular*}{\textwidth}{@{\extracolsep\fill}lll}
\toprule
\textbf{Name} & \textbf{Basis risk metric} & \textbf{Sources} \\
\midrule
Mean squared error (MSE) & $\mathbb{E}\left[ \bigl(S - Y\bigr)^2 \right]$ & \cite{Brown2007, Steinmann2023, Lemke2025} \\ \midrule
Relative residual variance\hyperlink{tfn1}{\tnote{1}} & $\frac{\Var\bigl(S - Y\bigr)}{\Var(S)}$ & \cite{Elabed2013, Stigler2023} \\  \midrule
Expected expenses & $ \sum\limits_{i,j =0}^1 C_{i,j} \mathbb{P}\left( \mathds{1}_{\{S > 0\}} = i, \mathds{1}_{\{Y > 0\}} = j \right)$, $C_{i,j} \ge 0$ & \cite{Figueiredo2018} \\ \midrule
Basis risk ratio & $\frac{\mathbb{E}[Y|S = s]}{m\mathbb{E}[S]}$, $s \in [0,L]$ =: Image(S), $L, m>0$ & \cite{Clarke2016} \\ \midrule
(Absolute relative) prediction error & $\mathbb{E}\left[\frac{|S-Y|}{S} \right]$ resp. $\mathbb{E}\left[S-Y\right]$ & \cite{Boyd2020} \\ \midrule
Basic / generalized penalty & $\mathbb{E}\left[f\bigl(S -Y \bigr) + \phi_n\bigl(S - Y \bigr) \right]$ resp. & \cite{Claramunt2022} \\ 
& $\mathbb{E}\left[ f\bigl(S - Y \bigr) +h_n \left(\frac{S - Y}{Y} \right)\phi_n\bigl(S - Y \bigr)\right]$; & \\
& $f(x) = \bigl(a_1 x + a_2 x^2\bigr)\mathds{1}_{\{x\le 0\}}$, $a_1 \le 0$, $a_2 \ge 0$, & \\
& $\phi_n(x) = \tilde{\phi}_n(x)\mathds{1}_{\{x\ge 0\}}$, & \\
& $ h_n: x \to \begin{cases}
h_n(x) = 1, \, & x \le r, \\
h_n(x) > 1, \, & x > r,
\end{cases}$ $\quad r>0$ &\\
& $\tilde{\phi}^{(k)}_n(x), h_n^{(k)}(x) \ge 0$, $\forall$ $x > 0$, $k=1,\ldots,n$.& \\
\bottomrule
\end{tabular*}
\begin{tablenotes}
\hypertarget{tfn1}{\item[1]} We note that minimization under this framework is equivalent to using the MSE basis risk metric if we restrict ourselves to payment schemes with a fixed mean, e.g., when demanding $\mathbb{E}[Y] = \mathbb{E}[S]$.
\end{tablenotes}
\end{threeparttable}
\end{table}
\par
Research on the optimal design of index insurance contracts can roughly be split into two categories: (i) focusing on customers' expected utility and (ii) direct measurement of basis risk, see \cite{Stigler2023}. Many contributions have been made under the first approach, evaluating underlying indices or the related question of trigger design (see, e.g., \cite{Vroege2021,Lichtenberg2022}) as well as payment schemes (see, e.g., \cite{Mahul1999, Mahul2001,Mahul2003, Bourgeon2003, Zhang2019, Louaas2024}) on varying levels of generality\footnote{Scopes reach from assessing specific improvements for a given index (e.g., enhancing a soil moisture index for parametric drought insurance through satellite data \cite{Vroege2021}) to providing optimality results for an arbitrary underlying parameter (see \cite{Louaas2024}).}. Research in the second category is typically concerned with minimizing an appropriate basis risk metric. Usually, these approaches limit their scope to a given risk scenario (e.g., water supply risk \cite{Brown2007}, natural hazards \cite{Steinmann2023}, agriculture risk \cite{Stigler2023}) and might further restrict themselves to only comparing specific contract designs (e.g., three different indices for crop yield risk \cite{Boyd2020}) to allow for the use of real-world data.
\par 
We aim at generalizing the above considerations of explicit basis risk reduction in parametric contracts by deriving an expression for the optimal compensation under a general index. We show that the optimal payment scheme is defined through conditional expectiles of the policyholder's true loss given payment being triggered under pure parametric insurance with binary payouts. For variable compensation under parametric index insurance, this generalizes to expectiles of the true loss conditioned on the concrete index realization. To the best of our knowledge, we are the first to provide guidance on the explicit construction of basis risk-optimal payment schemes for any given, generic index. This generality enables a combination of our results with existing insights on the appropriate selection of product-specific indices and trigger points, complementing prior research. Furthermore, using properties of expectiles, we can connect the optimal payment amount to notions of risk perception and measurement as well as derive monotonicity results under different stochastic orderings of policyholders' conditional true losses. From a practical point-of-view, our theoretically optimal compensation structure can easily be implemented and connected to commonly offered payment schemes (e.g., linear or step payout functions \cite{Steinmann2023}), underscoring the value of our results for the actuarial community. 
\par
This paper is structured as follows. Section~\ref{sec2} recalls definitions and properties of (conditional) expectiles. In Section~\ref{sec3}, we specify our minimization framework, accounting for aspects of basis risk considered relevant in the existing literature, and derive the basis risk-optimal payment scheme for a general index. To this end, following the categorization of index insurance from \cite{Lin2020}, we first consider pure parametric insurance's binary compensation structures in Section~\ref{subsec2} before generalizing to payout functions used in parametric index insurance in Section~\ref{subsec3}. To visualize our findings, we conduct a simulation study on a parametric cloud outage cover in Section~\ref{sec4}. Section~\ref{sec5} concludes and provides an outlook for future research. 

\section{Background on (conditional) expectiles}\label{sec2}

In preparation for the derivation of basis risk-optimal payment schemes, we recall important definitions and properties of expectiles, statistical functionals first introduced in \cite{Newey1987}, as well as of conditional expectiles, an extension by \cite{Bellini2018}. We assume all random variables (rv) to be defined on a probability space $\left(\Omega, \mathcal{F},\mathbb{P}\right)$. Further, we denote by $\mathcal{L}^p\left(\Omega, \mathcal{G},\mathbb{P}\right)$, $p \in \mathbb{N}$, $\mathcal{G} \subset \mathcal{F}$ a (sub-)$\sigma$-algebra, the set of $\mathcal{G}$-measurable rvs with finite $p$-th moment. Likewise, $\mathcal{L}^\infty \left(\Omega, \mathcal{G},\mathbb{P}\right)$ describes the set of all bounded $\mathcal{G}$-measurable rvs. For notational convenience, we abbreviate these expressions by $\mathcal{L}^p$, resp.\ $\mathcal{L}^\infty$, if $\mathcal{G} = \mathcal{F}$.
\par 
For a rv $X$, expectiles asymmetrically extend the concept of the mean $\mathbb{E}[X]$, similar to the relationship between quantiles and the median \cite{Daouia2024}. The $\gamma$\emph{-expectile} $e_\gamma(X)$ of an $\mathcal{L}^1$ rv $X$ with $\gamma\in(0,1)$ is defined as the unique solution of the minimization problem 
\begin{equation*}
\min_{y \in \mathbb{R}} \, \mathbb{E}\big[\eta_\gamma(X-y) - \eta_\gamma(X) \big],
\end{equation*}

\pagebreak \noindent
where $\eta_\gamma(x):=|\gamma-\mathds{1}_{\{x \le 0\}}|x^2$ and $\mathds{1}_{\{\cdot\}}$ denotes the indicator function \cite{Newey1987}. For any $X\in \mathcal{L}^2$, an equivalent definition of the $\gamma$-expectile, see \cite{Bellini2018}, that is used in Section~\ref{sec3}, is given by
\begin{equation}\label{DefExpectile}
e_\gamma(X) =\underset{y \in \mathbb{R}}{\mathrm{argmin}}\left( \gamma \mathbb{E}\left[\left((X-y)^+\right)^2 \right] + (1-\gamma) \mathbb{E}\left[\left((X-y)^-\right)^2 \right] \right),
\end{equation}
where $x^+ := \max\{x,0\}$ and $x^- := \max\{-x,0\}$ are the positive, resp.\ negative, part of $x \in \mathbb{R}$. Expectiles possess a number of appealing properties, particularly when used as risk measures. For the reader's convenience, we shortly summarize important aspects, starting with basic characteristics presented in \cite[Theorem 1]{Newey1987} and \cite[Theorem 1]{Bellini2016}.
\begin{proposition}[Properties of expectiles, see \cite{Newey1987, Bellini2016}] \label{Prop1}
For $\gamma\in (0,1)$ and $X,Y \in \mathcal{L}^1$, $X\sim F$, it holds:
\begin{enumerate}
\item $e_\gamma(X+h) = e_\gamma(X)+h$ for all $h \in \mathbb{R}$. \label{Prop1.1}
\item $e_\gamma(\lambda X) = \lambda e_\gamma(X)$ for all $\lambda > 0$. \label{Prop1.2}
\item If $\gamma \le 1/2$, then $e_\gamma(X+Y) \ge e_\gamma(X) + e_\gamma(Y)$. If $\gamma \ge 1/2$, then \label{Prop1.3}
\begin{equation*}
e_\gamma(X + Y) \le e_\gamma(X) + e_\gamma(Y).
\end{equation*}
\item $e_\gamma(X)$ is strictly increasing with respect to $\gamma$.
\item If $F$ is continuously differentiable, then so is $e_\gamma(X)$ and for $(x, \gamma)$ with $F(x) \in (0,1)$, $x \neq \mathbb{E}[X]$, and $x = e_\gamma(X)$, 
\begin{equation*}
F(x) = -\frac{x -  \mathbb{E}[X] +\gamma(1 - 2\gamma) \frac{\partial}{\partial \gamma} e_\gamma(X)}{(1 - 2\gamma)^2\frac{\partial}{\partial \gamma} e_\gamma(X)}.
\end{equation*}
This equation holds in the limit for $x = \mathbb{E}[X]$ and $\gamma = 1/2$.\label{Prop1.5}
\end{enumerate}
\end{proposition} 
It follows from Properties~\ref{Prop1.1}., \ref{Prop1.2}., and \ref{Prop1.3}.\ that $\mathcal{L}^1 \ni X \mapsto e_\gamma(X)$ defines a coherent risk measure for any $\gamma \ge 1/2$\footnote{Besides being elicitable and coherent, expectiles with $\gamma \ge 1/2$ are the only coherent shortfall risk measures, see \cite{Weber2006}, and the only coherent risk measures defined by generalized quantiles, see \cite{Bellini2014}.}. Further, it will prove useful to assess under which circumstances the $\gamma$-expectile of a given rv $Y$ dominates the $\gamma$-expectile of another rv $X$, i.e. $e_\gamma(X) \le e_\gamma(Y)$. For this, we utilize stochastic order relations\footnote{In the order of Definition~\ref{DefStoOrders}: First order stochastic dominance, convex order, $\kappa$-almost first order stochastic dominance.}, which we define in the following.

\begin{definition}[Stochastic order relations]\label{DefStoOrders}
Given $\kappa \in (0,1)$ and rvs $X$ and $Y$, we write
\begin{align*}
X \le_{st} Y & \Leftrightarrow \mathbb{E}[u(X)] \le \mathbb{E}[u(Y)], \forall u \in \{u:\mathbb{R} \to \mathbb{R}| \,u \, \mathrm{increasing}\}, \\ 
X \le_{cx} Y & \Leftrightarrow \mathbb{E}[u(X)] \le \mathbb{E}[u(Y)], \forall u \in \{u:\mathbb{R} \to \mathbb{R}| \,u \, \mathrm{convex}\}, \\ 
X \le_{\kappa} Y & \Leftrightarrow \mathbb{E}[u(X)] \le \mathbb{E}[u(Y)], \forall u \in \{u:\mathbb{R} \to \mathbb{R}| \, 0 \le \kappa u'(y) \le u'(x), \, x,y\in\mathbb{R}\},
\end{align*}
where we tacitly assumed the existence of the corresponding expectations.
\end{definition}

Using the above notation, we can compile results on the monotonicity of expectiles from the works of \cite{Bellini2014,Bellini2012,DeVecchi2024} in the following proposition.
\begin{proposition}[Monotonicity properties of expectiles, see \cite{Bellini2014,Bellini2012,DeVecchi2024}]\label{Prop2} 
Let $0 < \kappa \le 1$ and $X,Y\in \mathcal{L}^1$. Then, one has
\begin{enumerate}
\item $X \le Y$ $\mathbb{P}$-a.s. $\Rightarrow$ $e_\gamma(X) \le e_\gamma(Y)$ for each $\gamma \in (0,1)$. \label{Prop2.1}
\item $X \le Y$ $\mathbb{P}$-a.s. and $\mathbb{P}(X < Y) > 0$ $\Rightarrow$ $e_\gamma(X) < e_\gamma(Y)$ for each $\gamma \in (0,1)$. \label{Prop2.2}
\item $X \le_{st} Y$ $\Rightarrow$ $e_\gamma(X) \le e_\gamma(Y)$ for each $\gamma \in (0,1)$. \label{Prop2.3}
\item $X \le_{cx} Y$ $\Rightarrow$ $e_\gamma(X) \le e_\gamma(Y)$ for each $\gamma \in [1/2,1)$. \label{Prop2.4}
\item $X \le_{\kappa} Y$ $\Rightarrow$ $e_\gamma(X) \le e_\gamma(Y)$ for each $\gamma \in \left[ \frac{\kappa}{1 + \kappa},\frac{1}{1 + \kappa} \right]$. \label{Prop2.6}
\end{enumerate}
\end{proposition}

Finally, let us recall the connection of expectiles to the actuarial fair premium of stop-loss indemnity insurance contracts (i.e. the stop-loss transform) for random variables with equal mean. In this context, \cite[Theorem 12]{Bellini2016} showed that dominance of expectiles for all levels $\gamma \in (0,1)$ is equivalent to a certain ordering of the underlying random variables' stop-loss transforms left and right of the (shared) mean. This result, reiterated below in Theorem~\ref{StopLoss}, allows us to link parametric insurance's optimal payment schemes to indemnity coverage contracts on the conditional true loss in Section~\ref{subsec2}.

\begin{theorem}[Expectiles and stop-loss insurance, see \cite{Bellini2016}]\label{StopLoss}
Let $X,Y \in \mathcal{L}^1$ with $\mathbb{E}[X] = \mathbb{E}[Y] = \mu$. Further, let
\begin{align*}
&\mathrm{ess \, inf}\left(X \right) := \sup \left\{ x \in \mathbb{R}: \, \mathbb{P}\left(X - x \le 0 \right)=0\right\}, \\
&\mathrm{ess \, sup}\left( X \right) := \inf \left\{x \in \mathbb{R}: \, \mathbb{P}\left( X-x \ge 0\right)=0\right\}, \\
&m := \max \left\{\mathrm{ess \, inf}\left( X\right), \mathrm{ess \, inf}\left(Y\right) \right\}, \quad M := \min\left\{\mathrm{ess \, sup}\left( Y\right), \mathrm{ess \, sup}\left( Y\right) \right\},
\end{align*}
and $\pi_X (s) := \mathbb{E}\left[ (X-s)^+\right]$, $\pi_Y (s) := \mathbb{E}\left[ (Y-s)^+\right]$ define the actuarial fair premium of a stop-loss (indemnity) insurance contract on $X$, resp.\ $Y$, with deductible $s$. Then, $e_\gamma(X) \le e_\gamma(Y)$ for any $\gamma \in (0,1)$ if and only if $\pi_X(s) \ge \pi_Y(s)$ for each $ s \in (m,\mu)$ and $\pi_X(s) \le \pi_Y(s)$ for each $s \in (\mu,M)$. 
\end{theorem} 

In search of an optimal payment design for parametric index insurance, we need a generalization of (\ref{DefExpectile}) that encompasses not only constants, but also the more general case of $\mathcal{L}^2(\Omega,\mathcal{G},\mathbb{P})$ rvs for some sub $\sigma$-algebra $\mathcal{G}\subset\mathcal{F}$. Such an extension is available with the concept of \emph{conditional expectiles}, see \cite{Bellini2018}, which are defined as follows.
\begin{definition}[Conditional expectiles, see \cite{Bellini2018}]\label{DefCExp}
Let $\mathcal{F},\mathcal{G}$ with $\mathcal{G} \subset \mathcal{F}$ be $\sigma$-algebras on $\Omega$. For any $\gamma \in (0,1)$, the \emph{conditional expectile} $e_\gamma(X|\mathcal{G})$ of a rv $X \in \mathcal{L}^2(\Omega,\mathcal{F},\mathbb{P})$ is $\mathbb{P}$-a.s.\ uniquely given by 
\begin{equation}\label{DefCondExp}
e_\gamma(X|\mathcal{G}) = \underset{Y \in \mathcal{L}^2(\Omega,\mathcal{G},\mathbb{P})}{\mathrm{argmin}} \mathbb{E}\left[ \gamma \bigl( (X-Y)^+\bigr)^2 + (1-\gamma) \bigl((X-Y)^-\bigr)^2 \right].
\end{equation}
\end{definition}
There exists an equivalent definition of conditional expectiles that only requires $X\in \mathcal{L}^1$. As (\ref{DefCondExp}) is sufficient for our purpose, however, this alternative is omitted and the interested reader is refer to \cite[Definition 2]{Bellini2018}. Additionally, for the sake of notational convenience, if $\mathcal{G}=\sigma(\mathcal{E})$, resp.\ $\mathcal{G}=\sigma(f)$, is generated by some set $\mathcal{E}$ or some measurable function $f$, we write $e_\gamma(X|\mathcal{E})$, resp.\ $e_\gamma(X|f)$, for $e_\gamma(X|\mathcal{G})$. Especially, note that for $A\in\mathcal{F}$ with $\mathbb{P}(A) > 0$, $e_\gamma(X|A)$ is equal to the unconditional expectile $e_\gamma(X)$ defined with respect to the conditional probability measure $\tilde{\mathbb{P}}(\circ) := \mathbb{P}\left( \circ |A\right)$.
\par 
The next lemma collects relevant properties of conditional expectiles.  Special attention is paid to Property~\ref{RegCondDistr}., which specifies a way to translate the results of Propositions~\ref{Prop1} and \ref{Prop2} from the unconditional to the conditional setting by working with appropriate regular conditional distributions. Further, it follows directly from Properties~\ref{CExpRM1}.\ to \ref{CExpRM3}.\ that $e_\gamma(X|\mathcal{G})$ for any $\gamma \ge 1/2$ defines a conditional coherent risk measure\footnote{See \cite{Detlefsen2005} and mind the sign change, as it employs a financial instead of an actuarial point of view.}. 
\begin{lemma}[Properties of $e_\gamma(\circ|\mathcal{G})$, see \cite{Bellini2018}]\label{PropCExp}
Let $e_\gamma(\circ|\mathcal{G})$ be given as in Definition~\ref{DefCExp}. Then, for any $\gamma_1, \gamma_2 \in (0,1)$ and $X, Y \in \mathcal{L}^2(\Omega,\mathcal{F},\mathbb{P})$, it holds:  
\begin{enumerate}
\item $\gamma_1 \le \gamma_2 \Rightarrow e_{\gamma_1}(X|\mathcal{G}) \le e_{\gamma_2}(X|\mathcal{G})$.
\item $X \le Y$ $\mathbb{P}$-a.s. $\Rightarrow e_\gamma(X|\mathcal{G}) \le e_\gamma(Y|\mathcal{G})$.  \label{CExpAS}
\item $e_\gamma(X+H|\mathcal{G}) = e_\gamma(X|\mathcal{G}) + H$ for any $H \in \mathcal{L}^2(\Omega,\mathcal{G},\mathbb{P})$. \label{CExpRM1}
\item $e_\gamma(\Lambda\, X|\mathcal{G}) = \Lambda\, e_\gamma(X|\mathcal{G})$ for any non-negative $\Lambda \in \mathcal{L}^\infty(\Omega,\mathcal{G},\mathbb{P})$.
\item If $\gamma \le 1/2$, $e_\gamma(X + Y|\mathcal{G}) \ge e_\gamma(X|\mathcal{G}) + e_\gamma(Y|\mathcal{G})$. If $\gamma \ge 1/2$,  \label{CExpRM3}
\begin{equation*}
e_\gamma(X + Y|\mathcal{G}) \le e_\gamma(X|\mathcal{G}) + e_\gamma(Y|\mathcal{G}).
\end{equation*}
\item $e_\gamma(X|\mathcal{G})(\omega) = e_\gamma\bigl(F_\mathcal{G}(\circ,\omega)\bigr)$, $\omega \in \Omega$, where $F_\mathcal{G}(\circ,\omega)$ is a regular conditional distribution of $X$ on $\mathcal{G}$. \label{RegCondDistr}
\end{enumerate}
Here, all (in)equalities between random variables are meant to hold $\mathbb{P}$-almost surely. Note that the above statements do in fact hold for $X,Y \in \mathcal{L}^1(\Omega,\mathcal{F},\mathbb{P})$ and $H \in \mathcal{L}^1(\Omega,\mathcal{G},\mathbb{P})$ when using $e_\gamma(\circ|\mathcal{G})$ as defined in \cite[Definition 2]{Bellini2018}. 
\end{lemma}
To conclude, let us briefly comment on the task of actually calculating (conditional) expectiles, following the excellent treatment in \cite{Daouia2024} and the references therein. While explicit expressions for expectiles of discrete distributions can always be derived, closed-form solutions to Equation~(\ref{DefExpectile}) are not available for most continuous distributions. As the latter are arguably of greater importance in the actuarial domain, this issue needs to be kept in mind when introducing expectile-based payment schemes. As the unconditional $\gamma$-expectile of an integrable $X\sim F$ is equal to the unique $x$ solving
\begin{equation}\label{ExpectileEq} 
\gamma = \frac{G(x) - xF(x)}{2\bigl(G(x) - xF(x)\bigr) + x - E}, 
\end{equation}   
with expectation $E=\int_{-\infty}^{\infty} t \, dF(t)$ and lower partial moment function $G(x) = \int_{-\infty}^{x} t \, dF(t)$, bisection search \cite{OttoSobotka2024} or Newton--Raphson and Monte Carlo algorithms \cite{Daouia2024} provide suitable approximations. Implementations in the software \texttt{R} are available in the packages \texttt{expectreg} \cite{OttoSobotka2024}, resp.\ \texttt{Expectrem} \cite{Daouia2024}. For deriving conditional expectiles, referring back to their roots in asymmetric least-squares regression \cite{Newey1987} provides an intuitive approximation tool embedded in actuarial practice (see also Sections~\ref{subsec2} and \ref{subsec3}). The package \texttt{expectreg} offers a user-friendly implementation in \texttt{R} via the function \texttt{expectreg.ls(...)}.

\section{Minimizing basis risk}\label{sec3}
We consider a portfolio of $K$ parametric insurance contracts, all written on the \emph{same} underlying index $\theta$ and covering the period $[0,T]$. The payout received by policyholder $k$ for the $i$-th incident at $t_i\in[0,T]$ is given by $Y_i^{(k)} = g^{(k)}(\theta_i, t_i)$ and depends on the deterministic payment scheme $g^{(k)}: \theta(\Omega) \times [0,T] \to \mathbb{R}_{\ge 0}$, the time $t_i$ at which the incident occurred, and the corresponding index measurement $\theta_i$. Thus, the compensation for policyholder $k$ following incident $i$ is a realization of the rv $Y^{(k)}:=g^{(k)}(\theta,\tau)$, where $\tau$ denotes the incident time. Both the policyholder's payment scheme $g^{(k)}$ and true loss $S^{(k)}$ may be influenced by covariates $\boldsymbol{x}^{(k)} = \bigl( x_1^{(k)}, \ldots, x_l^{(k)} \bigr)$, $l\in \mathbb{N}$, which could encapsulate policyholder characteristics (e.g., geographic location) and other influential factors.
\par 
To explore the design of basis risk-optimal payment schemes, we first introduce a formal definition of basis risk in Section~\ref{subsec1}. Building upon this setting, we derive the optimal fixed payment for a binary pure parametric insurance contract in Section~\ref{subsec2} before generalizing this result to generic payment schemes offered under parametric index insurance coverage in Section~\ref{subsec3}.

\subsection{A quantification framework for basis risk}\label{subsec1}
We aim at deriving basis risk-optimal payment schemes in a minimization context. Hence, we first provide a suitable quantification framework for the basis risk associated with policyholder $k$'s index insurance contract. Based on existing literature and general strategic considerations, we include the following characteristics:
\begin{enumerate}
\renewcommand{\labelenumi}{(\roman{enumi})}
\item Basis risk refers to a situation where $Y^{(k)}$, received by policyholder $k$, differs from the true loss $S^{(k)}$ they experienced. Recall from Table~\ref{BasisRiskMeasures} that the existing literature focuses on correlation and distance-based measures. Given that a high correlation between $Y^{(k)}$ and $S^{(k)}$ only indicates strong linear dependence, which is not identical to closeness between damage and compensation, particularly for large claims \cite{Lopez2022}, for our application the distance between $S^{(k)}$ and $Y^{(k)}$ provides a natural starting point.
\item Basis risk affects insurers and insureds in an adverse way, see \cite{Lin2020}. Positive basis risk ($Y^{(k)} > S^{(k)}$) unnecessarily increases premium amounts \cite{Figueiredo2018}, e.g., through safety margins \cite{Claramunt2022}, and impacts insurers' insolvency risk \cite{Lin2020}, while negative basis risk ($Y^{(k)} < S^{(k)}$) leaves policyholders under-protected and may lower renewal rates \cite{Lin2020}, potentially endangering long-term viability of the product. Thus, deviations in both directions should be penalized. 
\item Different actors might perceive basis risk arising from a certain difference between losses and payouts differently. While larger deviations should rationally be associated with higher (or at the very least equal) risk, the marginal increase in perceived risk can vary.
\end{enumerate}

Based on these considerations, we choose to define basis risk as follows:

\begin{definition}[Basis risk]
The perceived basis risk, arising from the individual parametric coverage contract between policyholder $k$ and the insurer, is given by
\begin{equation}\label{BRpure}
B^{(k)} = w^{(k)}\left( \alpha^{(k)} \bigl(S^{(k)} - Y^{(k)} \bigr)^+ +\bigl(1 - \alpha^{(k)}\bigr) \bigl(S^{(k)} - Y^{(k)} \bigr)^-  \right),
\end{equation}
where $S^{(k)}$ and $Y^{(k)}$ are non-negative\footnote{Note that we explicitly allow $\mathbb{P}(S^{(k)} = 0) > 0$ in contrast to indemnity insurance, where incidents necessarily coincide with strictly positive losses.} rvs representing the actual loss of, resp.\ the payout to, policyholder $k$. $\alpha^{(k)} \in (0,1)$ weighs the subjective importance of negative basis risk relative to positive basis risk. $w^{(k)}: \mathbb{R}_{\ge 0} \to \mathbb{R}_{\ge 0}$ is a continuous, non-decreasing function which describes the relationship between the actual deviation of losses from payment and the associated perceived basis risk.  
\end{definition} 
In the following, we aim at minimizing (\ref{BRpure}) with respect to $Y^{(k)} = g^{(k)}(\theta,\tau)$ under the assumption $w^{(k)}(x) = x^2$. With this choice\footnote{Note that $\bigl(S^{(k)} - Y^{(k)} \bigr)^+ \bigl(S^{(k)} - Y^{(k)} \bigr)^- = 0$.}, (\ref{BRpure}) simplifies to 
\begin{equation}\label{BRpureAssum}
B^{(k)} =  \left(\alpha^{(k)}\right)^2 \Bigl(\bigl(S^{(k)} - Y^{(k)} \bigr)^+\Bigr)^2 + \left(1-\alpha^{(k)}\right)^2  \Bigl(\bigl(S^{(k)} - Y^{(k)} \bigr)^-\Bigr)^2.
\end{equation}
This ansatz has appealing properties. First, it allows to embed our minimization problem in the realm of expectiles, providing explicit expressions for optimal payment schemes. Second, it generalizes the symmetric (root) mean square error optimization criterion suggested in, e.g., \cite{Brown2007,Steinmann2023}, while also following the idea of assessing basis risk through \emph{basic penalty functions}\footnote{$B^{(k)} = f\left(S^{(k)} - Y^{(k)} \right)\mathds{1}_{\{S^{(k)} - Y^{(k)} \le 0\}} + \Phi_n \left(S^{(k)} - Y^{(k)} \right)\mathds{1}_{\{S^{(k)} - Y^{(k)} \ge 0\}}$, with $\Phi_n (x) = \left(\alpha^{(k)} \right)^2 x^2$, $f(x) = \left(1-\alpha^{(k)} \right)^2 x^2$.}, see \cite[Definition 4.1]{Claramunt2022}, firmly anchoring our framework in existing research. Further, as will be shown in Propositions~\ref{OptBRpure} and \ref{OptBRindex}, the chosen risk model translates to an intuitive optimal contract design. Third, the quadratic structure of $w^{(k)}$ assigns a higher relative importance to larger deviations, which are more likely to endanger the long-term viability of a parametric product by leaving policyholders severely under-protected (eroding trust) or heavily affecting insurer liquidity. 
\par
Note that $S^{(k)},Y^{(k)} \in \mathcal{L}^2$ is implicitly assumed to guarantee well-definedness of (\ref{BRpureAssum}). For the actual loss $S^{(k)}$, this has to be considered in the underwriting process, as heavy-tailed losses should in general not be covered exclusively through parametric insurance, see \cite{Lopez2022}. Here, index insurance should be used as a complement to indemnity insurance \cite{SwissRe2023}, e.g., via an insurance tower. As the random payments $Y^{(k)}$ under any potential payment scheme $g^{(k)}$ should match $S^{(k)}$ reasonably well, the restriction to square-integrable compensation follows.

\subsection{Pure parametric insurance}\label{subsec2}
We start with the simple setting of pure parametric insurance, i.e.\ each policyholder's contract has a binary payment structure given by
\begin{equation} \label{PSpure}
g^{(k)}(\theta,\tau) = y^{(k)}\, \mathds{1}_{\{\theta \in \Theta^{(k)}\}},
\end{equation}
with fixed\footnote{(\ref{PSpure}) can be extended to allow for time-dependence by using $y_{var}^{(k)}: [0,T] \to \mathbb{R}_{\ge 0}$, resp. $\Theta^{(k)}_{var}: [0,T] \to \mathcal{P}\bigl(\theta(\Omega)\bigr)$. An example is provided in Section~\ref{sec4}.} payout $y^{(k)} \in \mathbb{R}_{> 0}$, received once the index is triggered, and \emph{trigger area} $\Theta^{(k)}$, containing all parameter values leading to a payment. In this setting, minimizing expected basis risk on individual policyholder level corresponds to selecting parameters $\left ( y^{(k)}, \Theta^{(k)}\right)$ that best reduce $\mathbb{E}\left[B^{(k)}\right]$ given by
\begin{equation}\label{ExpBRpure} 
\mathbb{E}\left[\left(\alpha^{(k)}\right)^2 \Bigl(\bigl(S^{(k)} - y^{(k)}\mathds{1}_{\{\theta \in \Theta^{(k)}\}} \bigr)^+\Bigr)^2 + \left(1-\alpha^{(k)}\right)^2  \Bigl(\bigl(S^{(k)} - y^{(k)}\mathds{1}_{\{\theta \in \Theta^{(k)}\}} \bigr)^-\Bigr)^2  \right] .
\end{equation}
The trigger area $\Theta^{(k)}$ is typically specified by practical considerations like the accuracy of index measurement or a preference for simple rules, e.g., $\Theta^{(k)} := \left(\vartheta^{(k)},\infty\right)$ for a fixed $\vartheta^{(k)}$. Hence, we consider $\Theta^{(k)}$ as given and focus on optimizing the payout $y^{(k)}$. The unique solution to this problem is the expectile of the actual loss $S^{(k)}$ conditioned on the payment being triggered, the expectile's level depending on the relative importance of negative basis risk. If positive and negative deviations are weighted equally, (scaled) minimal expected basis risk can be split into the two separate scenarios under which basis risk arises, weighted by their respective probability of occurrence: (i) Payout is triggered ($\theta \in \Theta$) but does not match the actual damages due to random deviation from the expected loss amount. (ii) No payment is triggered ($\theta \notin \Theta$), even though the policyholder suffers a loss. These results are formalized below, a proof is given in Appendix~\ref{ProofPureParametric}.
\begin{proposition}[Minimizing basis risk for pure parametric insurance] \label{OptBRpure}
Let $k\in\{1,\ldots,K\}$ and $\mathbb{E}\left[B^{(k)}\right]$ as in (\ref{ExpBRpure}). Then, for given $\Theta^{(k)} \subset \theta(\Omega)$ with $\mathbb{P}(\theta \in \Theta^{(k)}) > 0$,
\begin{align}\label{proof1}
&e_ {\gamma^{(k)}}\left(S^{(k)} \middle| \theta \in \Theta^{(k)}\right) =\underset{y^{(k)} > 0}{\mathrm{argmin}} \, \mathbb{E}\left[B^{(k)}\right],\; \gamma^{(k)} = \frac{\big( \alpha^{(k)} \big)^2}{\big(1 - \alpha^{(k)}\big)^2 + ( \alpha^{(k)} )^2} \in (0,1),
\end{align}
uniquely. Especially, if $\alpha^{(k)} = 0.5$,
\begin{align}\label{proof2}
&\min_{y^{(k)} > 0} \mathbb{E}\left[B^{(k)}\right] = \nonumber \\
 &\qquad \frac{1}{4} \Biggl( \mathbb{P}\left(\theta \in \Theta^{(k)}\right) \Var\left(S^{(k)}\middle|\theta \in \Theta^{(k)} \right)+ \mathbb{P}\left(\theta \notin \Theta^{(k)}\right) \mathbb{E}\left[ \bigl(S^{(k)}\bigr)^2 \middle| \theta \notin \Theta^{(k)} \right] \Biggr).
\end{align}
\end{proposition}
Proposition~\ref{OptBRpure} implicitly assumes $e_{\gamma^{(k)}} \left(S^{(k)}| \theta \in \Theta^{(k)}\right) > 0$, which immediately follows from Property~\ref{Prop2.2} in Proposition~\ref{Prop2}, as the insurable loss $S^{(k)}$ is non-negative and no parametric coverage would reasonably be offered for a trigger area where $\mathbb{P}(S^{(k)} = 0| \theta \in \Theta^{(k)}) = 1$\footnote{Clearly, $y = 0$ is the (unique) minimizer of $\gamma^{(k)}\mathbb{E}\left[ \bigl((0 -y)^+\bigr)^2 \big| \theta \in \Theta\right] + (1 - \gamma^{(k)})\mathbb{E}\left[ \bigl((0 -y)^-\bigr)^2 \big| \theta \in \Theta\right]$.}.
\par 
\begin{remark}[Properties of the optimal solution]
Properties of expectiles translate to an intuitive behavior of the optimal payment scheme (\ref{proof1}). First, as $\gamma^{(k)}$ is strictly increasing in $\alpha^{(k)}$, increasing the weight (and thus penalty) of negative basis risk results in a larger optimal fixed payout. More precisely, Propositions~\ref{Prop1} and \ref{OptBRpure} allow to explicitly quantify the increase in payment associated with a shift in weight from positive to negative basis risk. If the conditional distribution of $S^{(k)}|\theta \in \Theta^{(k)}$ is continuously differentiable, we can quantify the \emph{marginal} change in optimal compensation via Property~\ref{Prop1.5} in Proposition~\ref{Prop1}. Second, in a realistic setting, we expect $\alpha^{(k)} \ge 1/2$ ($\Leftrightarrow \gamma^{(k)} \ge 1/2$), as policyholders are less sensitive to positive basis risk that primarily impacts the insurer in the short term \cite{Lin2020}. With that, the optimal payout (\ref{proof1}) to policyholder $k$ corresponds to the risk associated with their true loss $S^{(k)}$ under the elicitable, coherent risk measure $e_{\gamma^{(k)}}\left(\circ\middle| \theta \in \Theta^{(k)}\right)$. Third, we can relate the magnitude of payouts to policyholders $k_1$ and $k_2$, sharing the relative risk weighting $\alpha^{(k_1)} = \alpha^{(k_2)}$ and trigger area $\Theta^{(k_1)} = \Theta^{(k_2)}$, to the stochastic ordering of their actual losses conditioned on trigger requirements being met. Specifically, the optimal payment to policyholder $k_2$ dominates the respective compensation for policyholder $k_1$:
\pagebreak
\begin{enumerate}
\renewcommand{\labelenumi}{(\roman{enumi})}
\item (strictly) for any $\alpha^{(k_1)}$, if $S^{(k_2)}$ (strictly) dominates $S^{(k_1)}$ $\mathbb{P}\left(\circ \middle|\theta \in \Theta^{(k_1)}\right)$-a.s.
\item for any $\alpha^{(k_1)}$, if $S^{(k_2)}$ dominates $S^{(k_1)}$ in first order under $\mathbb{P}\left(\circ \middle|\theta \in \Theta^{(k_1)}\right)$.
\item for any $\alpha^{(k_1)} \ge 1/2$, if $S^{(k_2)}$ dominates $S^{(k_1)}$ in convex order under $\mathbb{P}\left(\circ \middle|\theta \in \Theta^{(k_1)}\right)$.
\item for any $\alpha^{(k_1)} \in \left[\frac{\sqrt{\kappa} - \kappa}{1-\kappa}, \frac{1 - \sqrt{\kappa}}{1-\kappa} \right]$, if $S^{(k_2)}$ dominates $S^{(k_1)}$ in $\kappa$-almost first order under $\mathbb{P}\left(\circ \middle|\theta \in \Theta^{(k_1)}\right)$\footnote{The bonds on $\gamma^{(k)}$ given in Proposition~\ref{Prop2} can be easily translated to bonds on $\alpha^{(k)}$ using monotonicity and trivial arithmetic.}.
\end{enumerate}  
Additionally, we can link the ordering of optimal payments to the above policyholders $k_1$ and $k_2$ to the actuarial fair premia of stop-loss indemnity insurance contracts on policyholders' true loss given payment being triggered $\left(S^{(k)}|\theta \in \Theta^{(k)}\right)$, provided they suffer the same average damage in a triggering incident. The proof follows directly by replacing $\mathbb{P}$ in Theorem~\ref{StopLoss} with $\mathbb{P}\left(\circ\middle | \theta = \Theta^{(k_1)}\right)$.
\end{remark}
\begin{corollary}
Let $k_1,k_2 \in \{1,...,K\}$, $\Theta^{(k_1)} = \Theta^{(k_2)}$, and $\mathbb{E}\left[ S^{(k_1)} \middle|\theta \in \Theta^{(k_1)}\right] = \mu = \mathbb{E}\left[ S^{(k_2)} \middle|\theta \in \Theta^{(k_2)}\right]$. Further, for $ k=k_1,k_2$, let
\begin{align*}
&\mathrm{ess \, inf}\left( S^{(k)} \middle|\theta \in \Theta^{(k)}\right) := \sup \left\{ s \in \mathbb{R}: \, \mathbb{P}\left( S^{(k)} - s \le 0 \middle|\theta \in \Theta^{(k)}\right)=0\right\}, \\
&\mathrm{ess \, sup}\left( S^{(k)} \middle|\theta \in \Theta^{(k)}\right) := \inf \left\{ s \in \mathbb{R}: \, \mathbb{P}\left( S^{(k)} - s \ge 0 \middle|\theta \in \Theta^{(k)}\right)=0\right\}, \\
&m_{(\Theta,k_1,k_2)} := \max \left\{\mathrm{ess \, inf}\left( S^{(k_1)} \middle|\theta \in \Theta^{(k_1)}\right), \mathrm{ess \, inf}\left( S^{(k_2)} \middle|\theta \in \Theta^{(k_2)}\right) \right\},\\
&M_{(\Theta,k_1,k_2)} := \min\left\{\mathrm{ess \, sup}\left( S^{(k_1)} \middle|\theta \in \Theta^{(k_1)}\right), \mathrm{ess \, sup}\left( S^{(k_2)} \middle|\theta \in \Theta^{(k_2)}\right) \right\}, \\
&\pi^{(k)}(s) := \mathbb{E}\left[ \bigl(S^{(k)} - s\bigr)^+\middle| \theta \in \Theta^{(k)}\right].
\end{align*}
Then, the optimal fixed payment (\ref{proof1}) to policyholder $k_2$ dominates the respective compensation for policyholder $k_1$ for any $\alpha^{(k_1)} = \alpha^{(k_2)}$ if and only if $\pi^{(k_1)}(s) \ge \pi^{(k_2)}(s)$, for each $s \in \left(m_{(\Theta,k_1,k_2)},\mu\right)$, and $\pi^{(k_1)}(s) \le \pi^{(k_2)}(s)$, for each $s \in \left(\mu,M_{(\Theta,k_1,k_2)}\right)$. 
\end{corollary} 
\begin{remark}[Real-world implementation via asymmetric least squares regression]
In practice, the optimal payment scheme from Proposition~\ref{OptBRpure} can be approximated through an asymmetric least squares regression. While there exist more sophisticated approaches for estimating unconditional expectiles, as detailed at the end of Section~\ref{sec2}, their application to deriving (\ref{proof1}) requires knowledge of the distribution of $S^{(k)}|\theta \in \Theta^{(k)}$, which in some cases is difficult to find. Further, a regression framework allows for the implementation and interpretation of relevant covariate effects. Thus, following \cite{Newey1987}, the optimal fixed payment $\hat{e}_{\gamma^{(k)}}\left(S^{(k)} \middle|\theta \in \Theta^{(k)} \right)=\beta_0(\boldsymbol{x}^{(k)}) + \beta_1(\boldsymbol{x}^{(k)})$ is estimated by minimizing the (asymmetric) loss 
\begin{alignat}{2} \label{PPIExpRegr} 
L_{\gamma^{(k)} }\left( \boldsymbol{s}^{(k)}, \boldsymbol{\theta}^{hist} \right) :=& \sum_{m = 1}^{M} \gamma^{(k)} \mathds{1}_{\{s_m^{(k)} > \hat{s}_m^{(k)}\}}\left( s_m^{(k)} - \hat{s}_m^{(k)} \right)^2 + \nonumber \\
&\qquad \sum_{m = 1}^{M} \left( 1 - \gamma^{(k)} \right)\mathds{1}_{\{s_m^{(k)} < \hat{s}_m^{(k)}\}} \left( s_m^{(k)} - \hat{s}_m^{(k)} \right)^2,\\
\hat{s}_m^{(k)} :=& \beta_0(\boldsymbol{x}^{(k)}) + \beta_1(\boldsymbol{x}^{(k)})\mathds{1}_{\{ \theta_m^{hist} \in \Theta^{(k)}\}}, \quad \beta_i(\boldsymbol{x}^{(k)}) := \sum_j \beta_{ij}x^{(k)}_j, \nonumber
\end{alignat}
for parameters $\{\beta_{ij}\}_{(i,j) \in \{0,1\} \times  \{1,\ldots,l\}} \in \mathbb{R}^{2l}$ based on historical loss and index observations $\big( \boldsymbol{s}^{(k)}, \boldsymbol{\theta}^{hist} \big) = \big( s^{(k)}_m,  \theta_m^{hist} \big)_{m=1,\ldots,M}$. If a large number of varying expectile levels $\gamma^{(k)}$ needs to be considered, parametric expectile regression \cite{Gao2023} may be used to increase estimation efficiency.
\end{remark}
\subsection{Parametric index insurance}\label{subsec3}
We now lift our findings to more general parametric index insurance, where the realization of the parameter not only decides whether a payment is triggered but also determines its size, see \cite{Lin2020}. This gives rise to a variety of interesting payments schemes $g^{(k)}$, whose design should again be driven by a number of considerations. First, $g^{(k)}$ mimics the relationship between the underlying index and the covered losses, keeping basis risk low. Additionally, an appropriate payment scheme accounts for model complexity to reduce underwriting costs and to increase transparency for, and acceptance by, customers. This includes:
\begin{enumerate}
\renewcommand{\labelenumi}{(\roman{enumi})}
\item The payment scheme conforms to ``rational expectations''. Particularly, whenever $\theta$ measures the magnitude of a threat (e.g.\  wind speed, earth quake on Richter scale, etc.), $g^{(k)}$ should be increasing in $\theta$. Moreover, $g^{(k)}$ tends to be continuous w.r.t.\ $\theta$, so that small changes in index realization do not cause large jumps in the resulting payment.
\item The payment scheme ideally exhibits a common structure across all policyholders to allow for scalability. This also simplifies the statistical analysis in the context of risk modeling. One way to achieve this is specifying a parametric family for $g^{(k)}$ and then utilizing policyholder-appropriate parameters. The payment scheme (\ref{PSpure}) for pure parametric insurance is a special case, where the fixed payout $y^{(k)}$ and the trigger area $\Theta^{(k)}$ act as parameters.
\end{enumerate}
These requirements are accounted for rather naturally when estimating the optimal payment scheme for a given portfolio through regression. Thus, we allow any non-negative payment scheme $g^{(k)}$ for minimizing the expected basis risk of policyholder $k$, provided the resulting payments $Y^{(k)} = g^{(k)}(\theta,\tau)$ are sufficiently well-behaved to permit square-integrability, see Section~\ref{subsec1}. To additionally incorporate trigger areas $\Theta^{(k)}\subset \theta(\Omega)$, $\mathbb{P}\left(\theta \in \Theta^{(k)} \right) > 0$, any considered payment scheme must fulfill 
\begin{equation*}
g^{(k)}(\theta_i,t_i) = 0,\quad  \forall \theta_i \notin \Theta^{(k)}.
\end{equation*}
This is equivalent to generalizing the setting of Proposition~\ref{OptBRpure} by replacing the constant payment $y^{(k)}$ in (\ref{ExpBRpure}) with a non-negative, $\sigma(\theta,\tau)$-measurable random variable \mbox{$Y^{(k)} \in \mathcal{L}^2$}, i.e.\ our objective function $\mathbb{E}\big[B^{(k)}\big]$ is now given by
\begin{equation}\label{ExpBRindex} 
\mathbb{E}\left[\left(\alpha^{(k)}\right)^2 \Bigl(\bigl(S^{(k)} - Y^{(k)}\mathds{1}_{\{\theta \in \Theta^{(k)}\}} \bigr)^+\Bigr)^2 + \left(1-\alpha^{(k)}\right)^2  \Bigl(\bigl(S^{(k)} - Y^{(k)}\mathds{1}_{\{\theta \in \Theta^{(k)}\}} \bigr)^-\Bigr)^2  \right] .
\end{equation} 
Consequently, the optimal payment is now a conditional $\gamma^{(k)}$-expectile of the true loss $S^{(k)}$, which takes into account the actual incident time and index measurement. The scenario split of minimal expected basis risk under equal weighting of positive and negative deviations is extended in a similar fashion, see Appendix~\ref{ProofParametricIndex}.

\begin{proposition}[Minimizing basis risk for parametric index insurance]\label{OptBRindex}
Let $k\in\{1,\ldots,K\}$ and $\mathbb{E}\big[B^{(k)}\big]$ be given as in (\ref{ExpBRindex}). Then, the basis risk-optimal payment scheme $g^{(k)}_{opt}(\theta,\tau)$, with given trigger area $\Theta^{(k)} \subset \theta(\Omega)$ s.t.\ $\mathbb{P}\left(\theta \in \Theta^{(k)} \right) > 0$, is $\mathbb{P}$-a.s. uniquely given by
\begin{align}\label{proof1Index}
g^{(k)}_{opt}(\theta,\tau) &= \tilde{g}^{(k)}(\theta,\tau)\,\mathds{1}_{\{\theta \in \Theta^{(k)}\}}, \nonumber \\
\tilde{g}^{(k)}(\theta,\tau) &= e_{\gamma^{(k)}}\left(S^{(k)}\middle| \theta,\tau\right) = \underset{Y^{(k)} \in \mathcal{L}^2 (\Omega, \sigma(\theta,\tau), \mathbb{P}), Y^{(k)} \ge 0 \, \mathrm{a.s.}}{\mathrm{argmin}} \mathbb{E}\left[B^{(k)} \right], \nonumber\\
\gamma^{(k)} &= \frac{\big( \alpha^{(k)} \big)^2}{\big(1 - \alpha^{(k)}\big) ^2 + \big( \alpha^{(k)} \big)^2} \in (0,1),
\end{align}
in the sense that $\tilde{g}^{(k)}(\theta_i,t_i) = e_{\gamma^{(k)}}\left(S^{(k)}|\theta,\tau \right)(\omega)$ for some\footnote{The precise choice of $\omega$ does not matter, as $e_{\gamma^{(k)}}\big(S^{(k)}\big|\theta,\tau\big)$ is measurable w.r.t.\ $\sigma(\theta,\tau)$.} $\omega \in \{(\theta,\tau) = (\theta_i, t_i)\} \subset \Omega$, as defined in Lemma~\ref{PropCExp} (f).
Especially, if $\alpha^{(k)} = 0.5$,
\begin{align} \label{proof2Index}
&\min_{Y^{(k)} \in \mathcal{L}^2(\Omega, \sigma(\theta,\tau), \mathbb{P}), Y^{(k)} \ge 0 \, \mathrm{a.s.}} \mathbb{E}\left[B^{(k)} \right] =  \\
&\hspace{1.8cm} \frac{1}{4}\Biggl(\mathbb{P}\left(\theta \in \Theta^{(k)}\right)\mathbb{E}\left[\Var\left(S^{(k)} \middle| \theta, \tau \right) \middle| \theta \in \Theta^{(k)}\right] +\mathbb{P}\left(\theta \notin \Theta^{(k)}\right) \mathbb{E}\left[ \left( S^{(k)}\right)^2\middle| \theta \notin \Theta^{(k)}\right] \Biggr). \nonumber
\end{align}
\end{proposition}
Being a generalization of Proposition~\ref{OptBRpure}, the basis risk-optimal payment scheme (\ref{proof1Index}) shares structural similarities with the ones observed in Section~\ref{subsec2}. Many of the results on the relation to stochastic orderings of the underlying \emph{conditional} true losses, however, require hard-to-verify assumptions on the corresponding regular conditional distributions to allow the derivation via Proposition~\ref{Prop2} and Property~\ref{RegCondDistr} in Lemma~\ref{PropCExp}. Thus, we only comment on properties of (\ref{proof1Index}) that are likely to be applicable.
\begin{remark}[Properties of the optimal solution]
The basis risk-optimal payment scheme (\ref{proof1Index}) has the following properties:
\begin{enumerate}
\renewcommand{\labelenumi}{(\roman{enumi})}
\item It is increasing in the relative importance of negative basis risk $\alpha^{(k)}$.
\item It is monotone w.r.t.\ almost sure dominance, i.e.\ for policyholders $k_1,k_2$ with equal risk profile $\alpha^{(k_1)} = \alpha^{(k_2)}$ and trigger areas $\Theta^{(k_1)} = \Theta^{(k_2)}$, $S^{(k_1)} \le S^{(k_2)}$ $\mathbb{P}$-a.s.\ implies an a.s.\ non-decreasing optimal payment scheme for policyholder $k_2$, i.e., $g^{(k_2)}_{opt}(\theta,\tau) \ge g^{(k_1)}_{opt}(\theta,\tau)$.
\item For $\alpha^{(k)} \ge 1/2$, the optimal compensation $\tilde{g}^{(k)}(\theta,\tau)$ for policyholder $k$ is equal to the risk of their true loss under the conditional coherent risk measure $e_{\gamma^{(k)}}\left(\circ |\theta,\tau \right)$.
\end{enumerate}
\end{remark}
\begin{remark}[Real-world implementation via asymmetric least squares regression]
A real-world implementation is again available via an appropriate regression approach. This allows to assess the payment for an incident ($\theta_i, t_i$), overcoming a possible lack of information on the conditional loss distribution. Moreover, covariate effects can be implemented and interpreted. If an index can be measured with adequate precision, one can employ linear expectile regression, minimizing the (asymmetric) loss function:
\begin{align}\label{LinExpRegr}
\nonumber L_{\gamma^{(k)} }\left( \boldsymbol{s}^{(k)}, \boldsymbol{\theta}^{hist} \right) &:= \sum_{m = 1}^{M} \gamma^{(k)}\mathds{1}_{\{s_m^{(k)} > \hat{s}_m^{(k)}\}}\left( s_m^{(k)} - \hat{s}_m^{(k)} \right)^2 + \\
&\qquad \sum_{m = 1}^{M} \left( 1 - \gamma^{(k)} \right)\mathds{1}_{\{s_m^{(k)} < \hat{s}_m^{(k)}\}} \left( s_m^{(k)} - \hat{s}_m^{(k)} \right)^2, \nonumber \\
\hat{s}_m^{(k)} &:= \beta_0(\boldsymbol{x}^{(k)}) + \beta_1(\boldsymbol{x}^{(k)}) \theta_m^{hist}, \quad 
\beta_i(\boldsymbol{x}^{(k)}) := \sum_j \beta_{ij}x^{(k)}_j, \; \beta_{ij} \in \mathbb{R}
\end{align}
for historical loss and index observations $\big( \boldsymbol{s}^{(k)}, \boldsymbol{\theta}^{hist} \big) = \big( s^{(k)}_m,  \theta_m^{hist} \big)_{m=1,\ldots,M}$\footnote{While the above formula obviously assumes a static parametric contract, time dependence could easily be added to the model for $\hat{s}_m^{(k)}$ in the usual way.}. If index measurements are only available on a rough level, one could extend the conditional regression framework considered in (\ref{PPIExpRegr}). For example, if an insurer can only observe whether index values belong to one of the disjoint sets $\left \{ \Theta^{index}_1, \ldots, \Theta^{index}_Q \right \}$, an appropriate loss function for minimization is:
\begin{equation} \label{StepExpRegr}
\begin{split}
\nonumber L_{\gamma^{(k)} }\left( \boldsymbol{s}^{(k)}, \boldsymbol{\theta}^{hist} \right) &:= \sum_{m = 1}^{M} \gamma^{(k)}\mathds{1}_{\{s_m^{(k)} > \hat{s}_m^{(k)}\}}\left( s_m^{(k)} - \hat{s}_m^{(k)} \right)^2 + \\
&\qquad \sum_{m = 1}^{M} \left( 1 - \gamma^{(k)} \right)\mathds{1}_{\{s_m^{(k)} < \hat{s}_m^{(k)}\}} \left( s_m^{(k)} - \hat{s}_m^{(k)} \right)^2,\\
\hat{s}_m^{(k)} &:= \beta_0(\boldsymbol{x}^{(k)}) + \sum_{q=1}^Q \beta_q(\boldsymbol{x}^{(k)})\,\mathds{1}_{\{ \theta_m^{hist} \in \Theta^{index}_q\}}, 
\end{split}
\end{equation}
with $\big( \boldsymbol{s}^{(k)}, \boldsymbol{\theta}^{hist} \big)$ and $\beta_q(\boldsymbol{x}^{(k)})$, $q=0,\ldots,Q$, as above.
\par 
Using payment schemes like $g^{(k)}(\theta_i,t_i) = \beta_0(\boldsymbol{x}^{(k)}) + \beta_1(\boldsymbol{x}^{(k)}) \theta_i$ and $g^{(k)}(\theta_i,t_i) = \beta_0(\boldsymbol{x}^{(k)}) + \sum_{q=1}^Q \beta_q(\boldsymbol{x}^{(k)})\,\mathds{1}_{\{ \theta_i \in \Theta^{index}_q\}}$ is further supported by the practical considerations from the beginning of this section. These schemes are transparent to customers, can be scaled easily based on their common structure, and exhibit both continuity (if $\theta$ can be measured precisely enough for fitting (\ref{LinExpRegr})) and monotonicity, if these trends are present in sufficient strength. Additionally, the common use of linear, resulting from (\ref{LinExpRegr}), and step functions, resulting from (\ref{StepExpRegr}), as payment schemes in practice \cite{Steinmann2023} supports the use of these regression frameworks.
\end{remark}
Finally, note that the minimization approach has implicitly assumed that each loss event triggers a check for compensation, i.e.\ an incident, under parametric insurance. While generally unlikely, however, a ``loss without incident'' might be possible for certain indices, which gives rise to additional basis risk. We can account for this by replacing $Y^{(k)}$ in (\ref{BRpure}) by $Z^{(k)}\cdot Y^{(k)}$, with independent $Z^{(k)} \sim \text{Ber}\left(1-p^{(k)}\right)$ and $p^{(k)} \in (0,1)$ equal to the probability that policyholder $k$ suffers a ``loss without incident''. Simple tower property arguments, see Appendix~\ref{ProofLossWOIncident}, allow to prove small variations of Propositions~\ref{OptBRpure} and \ref{OptBRindex}, where the probability measure $\mathbb{P}$ underlying the expectile payment scheme is replaced by its conditional version given the loss coincides with an incident, i.e.\ $\mathbb{P}(\circ|Z^{(k)} = 1)$.  
\begin{corollary}[Optimal payment schemes in the presence of loss without incident] \label{LossWOIncident}
With $Z^{(k)}$ defined as above, it holds in the setting of Proposition~\ref{OptBRpure}:
\begin{equation*}
e_ {\gamma^{(k)}}\left(S^{(k)} \middle| \theta \in \Theta^{(k)};Z^{(k)} = 1\right) = \underset{y^{(k)} > 0}{\mathrm{argmin}}\,  \mathbb{E}\left[B^{(k)}\right].
\end{equation*}
In the generalized framework of Proposition~\ref{OptBRindex}, we have:
\begin{equation*}
e_{\gamma^{(k)}}\left(S^{(k)}|\theta,\tau; Z^{(k)} = 1\right) = \underset{Y^{(k)} \in \mathcal{L}^2 \bigl(\Omega, \sigma(\theta,\tau), \mathbb{P}\bigr), Y^{(k)} \ge 0 \, \mathrm{a.s.}}{\mathrm{argmin}} \mathbb{E}\left[B^{(k)} \right]. 
\end{equation*}
where $\gamma^{(k)}$ is defined as in Proposition~\ref{OptBRpure} and $e_{\gamma^{(k)}}(X|\mathcal{G}; Z^{(k)} = 1)$ denotes the conditional expectile of $X$ given $\mathcal{G}$ with respect to the conditional probability measure $\tilde{\mathbb{P}}(\circ) := \mathbb{P}(\circ|Z^{(k)} = 1)$, i.e.,
\begin{equation*}
e_{\gamma^{(k)}}(X|\mathcal{G}; Z^{(k)} = 1) =\underset{Y \in \mathcal{L}^2\bigl(\Omega,\mathcal{G},\tilde{\mathbb{P}} \bigr)}{\mathrm{argmin}} \mathbb{E}\left[ \gamma^{(k)} \bigl( (X-Y)^+\bigr)^2 + \bigl(1-\gamma^{(k)}\bigr) \bigl((X-Y)^-\bigr)^2 \right].
\end{equation*} 
\end{corollary}
While the above consideration is of theoretical interest, omitting losses without incident seems acceptable in practical applications. Firstly, because such scenarios are highly unlikely in a well-designed index, especially for events severe enough to cause a non-zero payment if they had been correctly identified as an incident. Second, reliably quantifying $p^{(k)}$ is very difficult if not impossible, so the benefits of this generalized model would likely be outweighed by the costs from increased complexity.
\subsection*{Aggregate loss index insurance as a first application}
We consider \emph{aggregate loss index insurance} as a first example, which -- although considered a distinct category of parametric insurance in \cite{Lin2020} -- can be interpreted as a special case of parametric index insurance. The difference to the investigation up to this point is the choice of index $\theta$, which in the present application is given as the aggregate loss in a predefined area or region, where policyholders' risks are assumed to be (relatively) homogeneous. Proposition~\ref{OptBRindex}, however, is unaffected by this choice of $\theta$, all results are effortlessly transferable. Therefore, we can interpret (\ref{proof1Index}) in the familiar context of aggregate loss index insurance against agriculture risks, usually referred to as \emph{area-yield crop insurance} \cite{Lin2020}.
\begin{example}[Optimal payment schemes in area-yield crop insurance]\label{AreaYieldExample}
Consider an area-yield crop insurance portfolio of $K=50$ farms with yearly crop yield $C^{(k)}$, $k\in\{1,\ldots,50\}$, and maturity $T=1$ year. At the end of the coverage period, policyholders are compensated if the index
\begin{equation*}
\theta = \sum_{k=1}^K C^{(k)},
\end{equation*}
i.e.\ the aggregate yield of all covered farms, falls short of a predefined threshold $\vartheta = 3000$, i.e.\ $\Theta^{(k)} = (-\infty,\vartheta)$ for each $k$, compare \cite{Stigler2023}. The true loss of farmer $k$ to be covered by the above contract is assumed to be a piecewise linear function of their crop yield $C^{(k)}$, given by
\begin{equation}\label{TrueLossAggII}
S^{(k)} = \begin{cases}
0, & C^{(k)} \ge c_{crit}, \\
c_{crit} - C^{(k)}, & C^{(k)} \in \left( c_{min}, c_{crit} \right), \\
c_{crit} - c_{min} =: s_{max}, & C^{(k)} \le c_{min},
\end{cases}
\end{equation}
see Figure~\ref{YIELDloss}. Thus, farmers do not experience any loss if their individual yield is above a critical level, say $c_{crit} = 40$. If $C^{(k)}$ falls below this threshold, losses grow linearly with decreasing yield until the minimum viable level $c_{min} = 10$, below which the year's harvest is assumed as a total loss. This bounds the damage by $s_{max}:= c_{crit} - c_{min}$.
\begin{figure}[t]
\centering
\begin{minipage}[t]{0.45\textwidth}
\includegraphics[width=\textwidth]{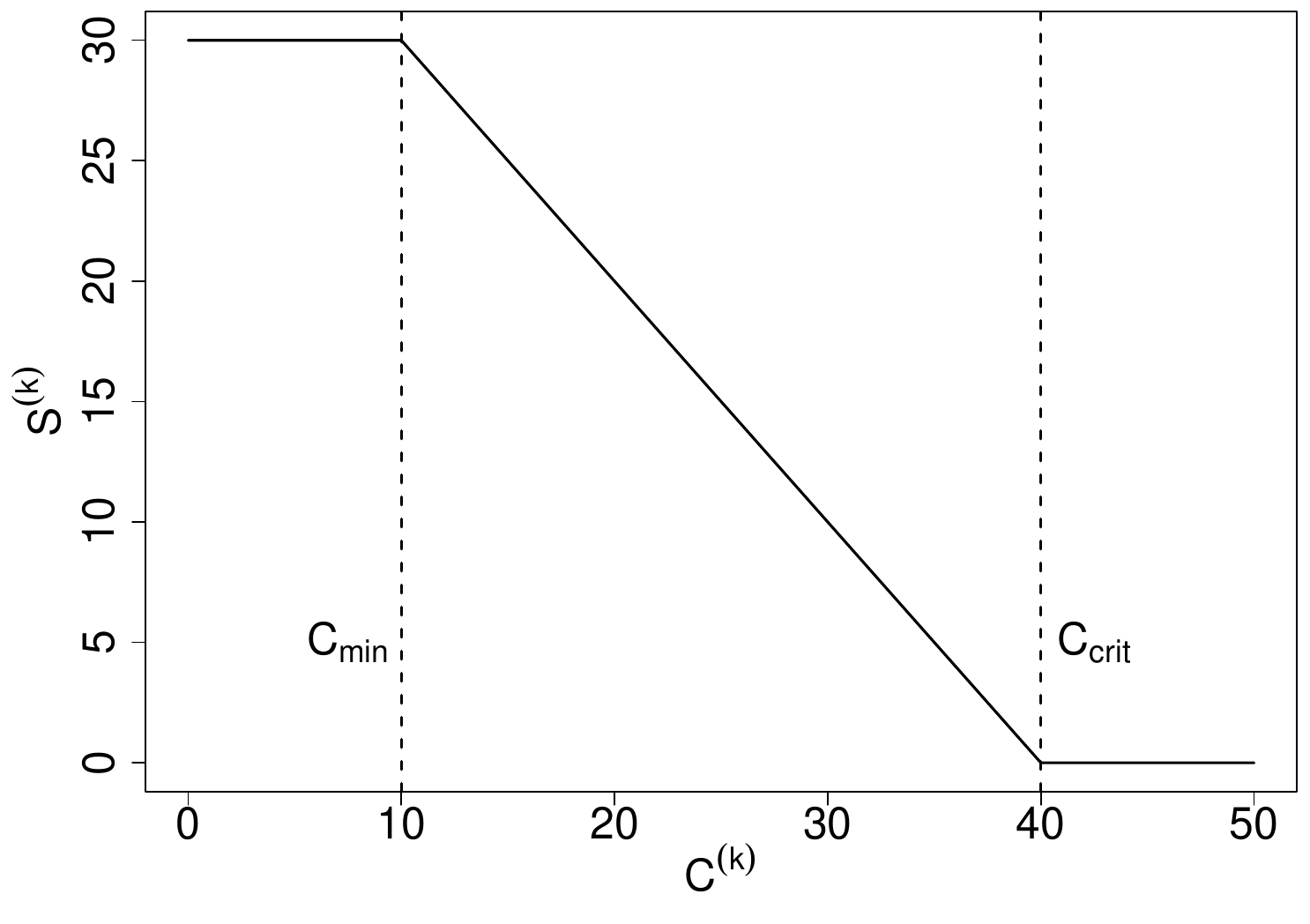}
\caption{True loss of policyholder $k$ for individual crop yield $C^{(k)}$.}\label{YIELDloss}
\end{minipage}
\hfill
\begin{minipage}[t]{0.45\textwidth}
\includegraphics[width=\textwidth]{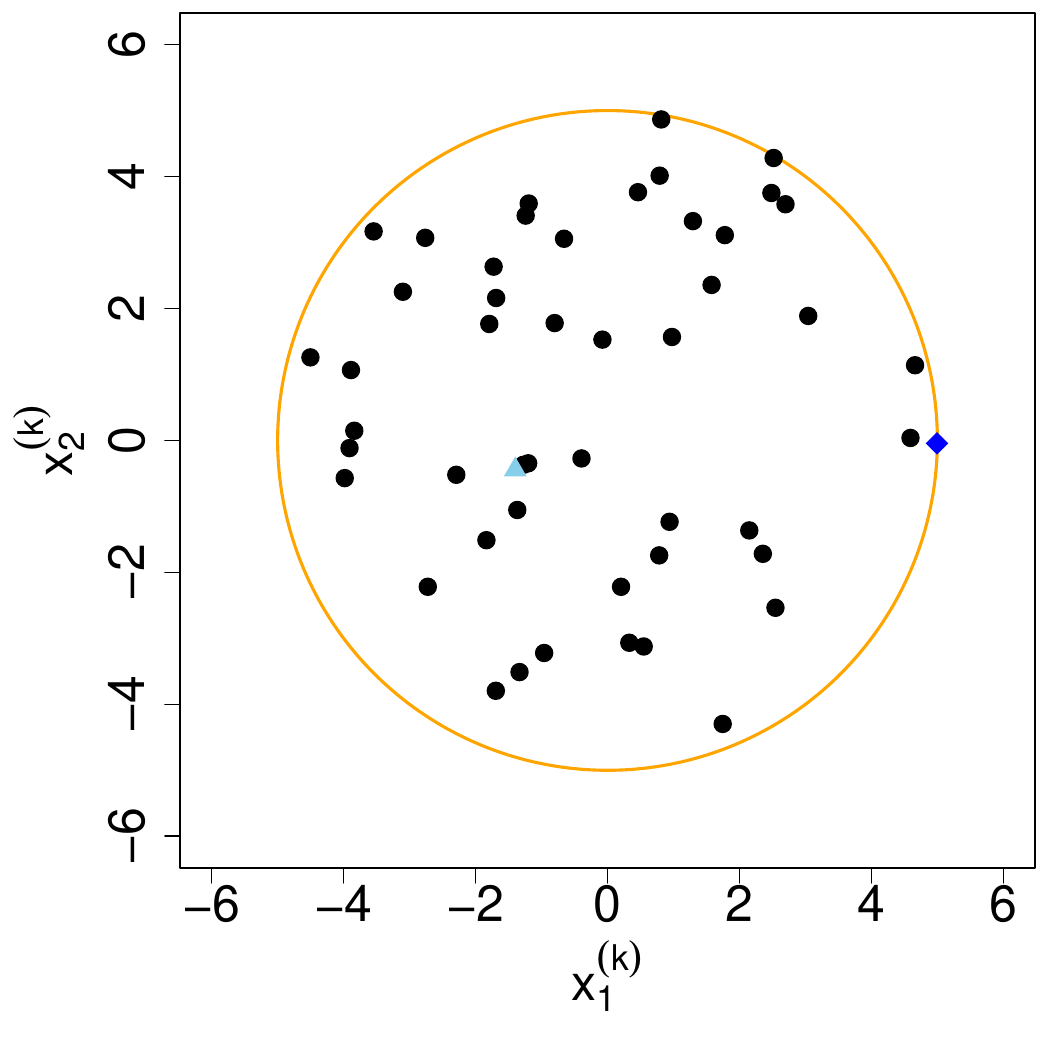}
\caption{Location of farms (black dots) in the covered area (orange circle). The two policyholders discussed in Figure~\ref{YIELDbasisrisk} are highlighted by a sky blue triangle $(-1.40,-0.44)$, resp. a blue diamond $(4.99,-0.05)$.}\label{YIELDportfolio}
\end{minipage}
\end{figure}

For modeling losses and payments, we assume farm locations 
\begin{equation*}
\bm{x}^{(k)} = \left(x^{(k)}_1, x^{(k)}_2\right) \in B_2(0,r):=\left \{\bm{x}\in \mathbb{R}^2 \, \middle| \, ||\bm{x}||_2 < r \right \}
\end{equation*}
to be uniformly distributed over the covered area; i.e.\ a circle with radius $r = 5$ around the origin, see Figure~\ref{YIELDportfolio}. Further, individual yields are assumed to be normally distributed\footnote{Normality of crop yields has been debated in literature and rejected for certain areas, see, e.g., \cite{Moss1993,Just1999,Ramirez2003}. As shown in \cite{Harri2009}, however, these non-normality results cannot be extrapolated to other regions and are crop-dependent. We have no stakes in this debate and assume normality to enjoy an educational example.} with mean $\mu^{(k)} \equiv \mu = 50$, standard deviation $\sigma^{(k)}  \equiv \sigma = 5$, and distance-dependent pairwise correlation
\begin{equation*}
\mathrm{Corr}\left(C^{(k)}, C^{(\ell)} \right) = \exp \left\{- \frac{||\bm{x}^{(k)} - \bm{x}^{(\ell)}||_2}{0.868r}\right\}, \; \forall k,\ell=1,\ldots,K.
\end{equation*}
Thus, the vector $\left(C^{(1)},\ldots, C^{(K)} \right)'$ arises from a two-dimensional Gaussian random field, see, e.g., \cite{Adler2007}, with exponential covariance \cite{Liu2019} evaluated at parameter $\bm{x} = \left(\bm{x}^{(1)}, \ldots, \bm{x}^{(K)}\right)$. Using standard properties of the multivariate normal distribution, we can easily derive the conditional distribution of policyholder $k$'s true loss given an observed index value of $\theta_i < \vartheta$, i.e.
\begin{align}
F_{S^{(k)}}(s) &:= \mathbb{P}\left(S^{(k)} \le s \middle| \theta = \theta_i \right) =\begin{cases}
0, & s < 0,\\
1 - \Phi_{\mu^*,\sigma^*}\left(c_{crit}-s\right), & s \in \left[0, s_{max}\right)\\
1, & s \ge s_{max},
\end{cases}\nonumber \\
\mu^* &= \mu + \frac{\boldsymbol{1}'\Sigma e_{[k]} }{\boldsymbol{1}'\Sigma \boldsymbol{1}} \left( \theta_i - K\mu \right) \in \mathbb{R},\quad  \label{YIELDcdfloss} \sigma^*=\sqrt{ \sigma^2 - \frac{\left( \boldsymbol{1}'\Sigma e_{[k]} \right)^2}{ \boldsymbol{1}'\Sigma \boldsymbol{1} } } \in \mathbb{R}_{>0}, \nonumber \\ 
\Sigma &= \Biggl( \sigma^2 \mathrm{Corr}\Bigl(C^{(i)}, C^{(j)} \Bigr) \Biggr)_{(i,j) \in \{1,\ldots,K\}^2} \in \mathbb{R}^{K\times K}, \nonumber\\
\boldsymbol{1} &= (1 \, \rule[0.5ex]{1.5em}{0.4pt} \,1)' \in \mathbb{R}^{K \times 1}, \nonumber \quad e_{[k]} = \bigl( \underbrace{0 \, \rule[0.5ex]{1.5em}{0.4pt} \, 0}_{\text{$k-1$ times}} 1 \underbrace{0 \, \rule[0.5ex]{1.5em}{0.4pt} \, 0}_{\text{$K-k$ times}} \bigr)' \in \mathbb{R}^{K \times 1}, \nonumber 
\end{align}
see Appendix~\ref{AppYieldLossDistr}. Here, $\Phi_{\mu,\sigma}(\cdot)$ denotes the cdf of an $\mathcal{N}\left(\mu,\sigma^2\right)$-distributed random variable. The optimal payment (\ref{proof1Index}) is found by numerically solving (\ref{ExpectileEq}), details are given in Appendix~\ref{AppYieldLossExpectile}.
\par 
The optimal payment scheme for two farms with approximate location $\bm{x}^{(14)}=(-1.40,-0.44)$, resp.\ $\bm{x}^{(16)}=(4.99,-0.05)$, is presented in Figure~\ref{YIELDbasisrisk}. As one expects from the theoretical findings, an increase in relevance of negative basis risk $\alpha^{(k)}$ results in higher optimal payments for any index realization $\theta_i$. Additionally, while payouts to both policyholders are growing with decreasing aggregated yield $\theta_i$, small index realizations have a far stronger impact on the compensation of the ``central'' policyholder with location $\bm{x}^{(14)}$. As the two considered policyholders were purposely chosen to exhibit maximum ($k=14$), resp. minimum ($k=16$), correlation of their individual yield $C^{(k)}$ with the index $\theta_i$ in our given portfolio\footnote{As $\sigma^{(k)}$ is equal across all policyholders, this is equivalent to maximizing/minimizing $\boldsymbol{1}'\Sigma e_{[k]}$.}, this is both intuitive and desirable, reflecting the severity of true losses we would expect given small aggregate yields.

\begin{figure}[h]
\center
\includegraphics[width = \textwidth, keepaspectratio]{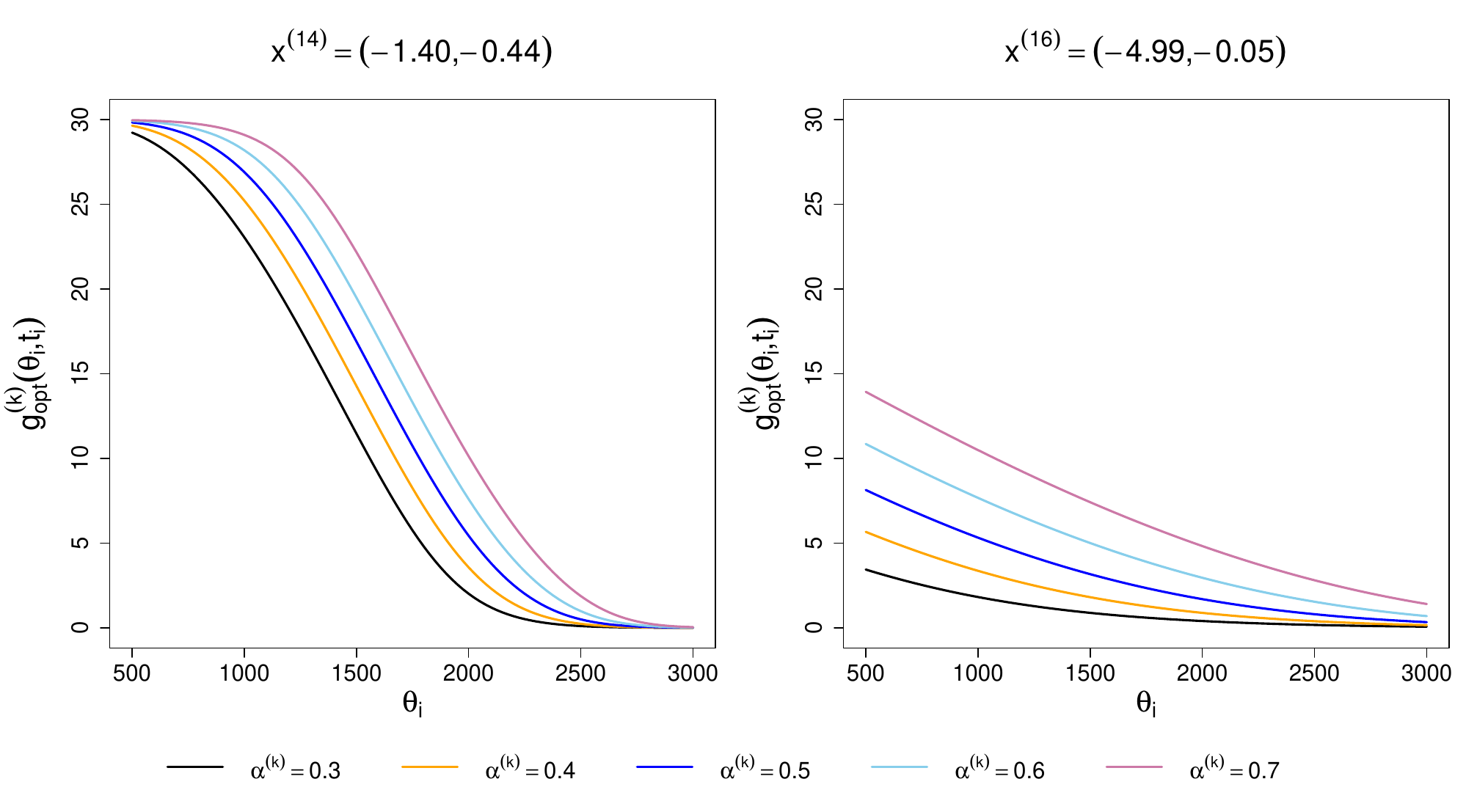}
\caption{Basis-risk optimal payment given the index realization $\theta_i \in [500,3000]$ to a policyholder with approximate location  $\bm{x}^{(14)} = (-1.40,-0.44)$ (left), resp.\ $\bm{x}^{(16)} = (4.99,-0.05)$ (right), for $\alpha^{(k)} \in \{0.3,0.4,0.5,0.6,0.7\}$.} 
\label{YIELDbasisrisk}
\end{figure}
\end{example}

\section{Parametric insurance for cloud outages}\label{sec4}
The optimal payment scheme and corresponding basis risk under pure parametric coverage are now applied to a specific cyber risk, i.e.\ we consider losses resulting from cloud outages. The rational behind the considered product is that off-times of cloud services can cause massive losses to companies relying on their services. Following a cloud downtime, affected companies may prefer a swift compensation over a long and difficult loss assessment in traditional indemnity insurance. Formalizing this, policyholder $k$ is compensated via the cloud outage index $\theta_{CO} = \big(\theta_{CO,(1)},\theta_{CO,(2)},\theta_{CO,(3)}\big)$, if at least one of three covered external service providers $j\in\{1,2,3\}$ suffers an outage whose duration $\theta_{CO,(j)}$ surpasses the lower threshold $\vartheta^{(k)}$. While index insurance is best studied for climate-related agriculture risks, see \cite{Louaas2024}, its use in cyber is considered in \cite{Lopez2022} and is further motivated by the 2020 launch of Lloyd's parametric cyber insurance against outage-related business interruptions, see \cite{Evans2020}. Data on cyber losses is scarce and seldom publicly available, see, e.g., \cite{Xu2018,Cremer2022,Zeller2022}. Hence, we design our study around a virtual portfolio, in which all distributional assumptions are motivated by existing research. 

\subsection{Baseline and simulation portfolio}\label{SimPort}

To generate loss and payment realizations, we conduct $N=10\,000$ simulation runs of outage events. We choose a yearly grid over the course of $T=5$ policy years and consider a fictitious portfolio with $K=500$ policyholders. Motivated by \cite{Zeller2022}, we combine $50$ policyholders with industry sectors $x_1^{(k)}=b^{(k)}$ and sizes $x_2^{(k)} = s^{(k)}$ into a \emph{baseline portfolio}, of which we consider $10$ copies with varying quantile levels $p^{(k)} \in \{0.05,0.1,\ldots,0.5\}$. The thresholds $\vartheta^{(k)}$ correspond to quantiles of the outage distributions, more details are provided in Section~\ref{TriggerPayment}. Table~\ref{SimParameters} presents an overview of chosen parameters.

\begin{table}[t]
\caption{Absolute and relative frequency of covariate levels in the portfolio as well as model areas affected by the respective characteristics. Extension of \cite[Table 4]{Zeller2022}.}\label{SimParameters}%
\begin{tabular*}{\textwidth}{@{\extracolsep\fill}llll}
\toprule
\textbf{Covariate} & \textbf{Scope} & \textbf{Relative (absolute)} & \textbf{Model area}\\
& & \textbf{frequency} & \textbf{affected}\\
\midrule
Sector $b^{(k)}$ & FI: Finance and insurance & 0.30 (150) & Fixed costs per outage\\
 & HC: Healthcare & 0.30 (150) & \\
 & BR: Businesses (retail) & 0.10 (50) & \\
 & EDU: Education & 0.10 (50) &\\
 & GOV: Government and military & 0.10 (50) &\\
 & MAN: Manufacturing & 0.10 (50) &\\ 
\midrule
Company & 1 Small & 0.60 (300) & Variable costs per hour\\
size $s^{(k)}$ & 2 Medium & 0.30 (150) & of outage duration\\
 & 3 Large & 0.10 (50) &\\ 
\midrule
Threshold & $\ell \cdot 0.05$, $\ell \in \{1,\ldots,10\}$ & 0.10 (50) & Trigger area\\
level $p^{(k)}$ & & constant for all $\ell$&\\
\bottomrule
\end{tabular*}
\end{table}

\subsection{Outage frequency}\label{Frequency}
We rely on Poisson processes with time-inhomogeneous rates $\lambda_{CO,(j)}(t) \ge 0$, $t\in [0,T]$, $j\in\{1,2,3\}$ to model the arrival of outage events, see \cite{Zeller2022}. Consequently, the number of outages experienced by service $j$ up to $t \in [0,T]$ is Poisson distributed, i.e.\
\begin{eqnarray}\label{FreqCO}
N_{CO,(j)}(t) &\sim & \text{Poi}\left(\int_0^t \lambda_{CO,(j)}(s) ds \right),\\
\lambda_{CO,(j)}(s) &=& \alpha_{CO,(j)}  s^{-\beta_{CO,(j)}},\label{RateWeibullPro}
\end{eqnarray}
the intensity follows the \emph{power law model}, a.k.a.\ \emph{Duane} or \emph{AMSAA model}, with parameters $\alpha_{CO,(j)} >0$ and $\beta_{CO,(j)} <1$. This has been suggested for modeling cloud service disruptions, see \cite{Tola2017}, and more generally for failures in repairable systems, see \cite{NISThandbook}. To illustrate differences in the behaviors over time, we choose the reliability developments detailed in \cite[Section 8.1.2.4]{NISThandbook}. Table~\ref{ParCOfreq} displays the implemented parameters, which are fitted to match the range of 4-year outage incident counts reported in \cite{Fiondella2013}, under the assumption that all services have the same expected number of outages $\alpha_{CO,(j)}(1 - \beta_{CO,(j)})^{-1}$ during the first year\footnote{We use the 4-year incident counts for  $\text{Heroku} = 39$, $\text{Amazon} = 17$, and $\text{Microsoft} = 10$. As $\mathbb{E}\left[N_{CO,(2)}(4)\right] = 4\alpha_{CO,(2)}$, we get $\alpha_{CO,(2)} = 4.25$ and by equality of first-year means $\frac{\alpha_{CO,(1)}}{1 - \beta_{CO,(1)}} = \frac{\alpha_{CO,(3)}}{1 - \beta_{CO,(3)}} = 4.25$. Thus, the remaining parameters are found by solving $39 = \mathbb{E}\left[N_{CO,(1)}(4)\right] =\frac{\alpha_{CO,(1)}}{1 - \beta_{CO,(1)}} 4^{1 - \beta_{CO,(1)}}$ resp. $10 = \mathbb{E}\left[N_{CO,(3)}(4)\right] =\frac{\alpha_{CO,(3)}}{1 - \beta_{CO,(3)}} 4^{1-\beta_{CO,(3)}}$ (rounded to two digits).}. 

Additionally, we make the following assumptions\footnote{For possible extensions, one can show that the time $T^{int}$ from current time $t_i$ until the next incident of a power law process with parameters $\alpha > 0$ and $\beta < 1$ has distribution
\begin{equation*}
\mathbb{P}\left( T^{int} \le x \right) = 1 - \exp\left( \frac{\alpha}{1-\beta} \bigl( (t_i+x)^{1-\beta} - t_i^{1-\beta} \bigr) \right), \, \forall x \ge 0. 
\end{equation*}} 
for modeling the occurrence of cloud outages:
\begin{enumerate}[start=1, label={(CO\arabic*)}]
\item Outages at different services arrive independent from one another, i.e.\ $\big(N_{CO,(j)}(t)\big)_{t \in [0,T]}$, $j\in\{1,2,3\}$ are independent. While in principle some dependence structure could be implemented, we refrain from doing so, lacking concrete information on a realistic dependence model for cloud outages.
\item The influence of outage duration on inter-arrival times is neglected. If a service has an outage at time $t_i$ with duration $\theta_i$, the earliest point in time at which a new outage is possible is $t_i + \theta_{i}$. As outage times are comparably small and rare, see (\ref{DurationModel}), allowing outages to take place in $(t_i, t_i + \theta_{i}]$ does not change our simulation results in a meaningful way.
\end{enumerate}
\begin{table}[h]
\begin{center}
\caption{Parameters chosen for the rates (\ref{RateWeibullPro}) of cloud services $j=1,2,3$.}\label{ParCOfreq}%
\begin{tabular}{@{}llrl@{}}
\toprule
\textbf{Service $j$} & $\boldsymbol{\alpha_{CO,(j)}}$  & $\boldsymbol{\beta_{CO,(j)}}$ &\textbf{Time trend}\\ 
\midrule
1 & $6.8$ & $-0.6$ & Deteriorating reliability \\
2 & $4.25$ & $0$ & Stable reliability \\
3 & $2.635$ & $0.38$ & Improving reliability \\ 
\bottomrule
\end{tabular}
\end{center}
\end{table}
\subsection{Outage duration and true losses}\label{Severity}
Policyholders' actual losses and compensation for the resulting service disruption will be driven by the outage duration $\theta_{CO,(j)}$, $j\in\{ 1,2,3\}$ on service-level. We make the following assumptions:
\begin{enumerate}[start=3, label={(CO\arabic*)}]
\item Durations $\theta_{CO,(j)}$ are independent and modeled on a yearly basis as identically distributed log-normal random variables, see \cite{Fiondella2013, Garraghan2014}. For ease of notation, we omit the subscript $(j)$ in the following.
\item Both parameters of the log-normal distributions increase in yearly, log-linear steps. The values are chosen such that the behavior in years $1$ and $5$ roughly resembles the outage durations observed for 2017 and 2022 in Uptime Intelligence's 2023 Annual outage analysis \cite{Uptime2023}. 
\end{enumerate}

In addition to the above setting, we consider an alternative, light-tailed duration model to compare the behavior of basis risk associated with our payment schemes under different tail weights. For this, we use a Gamma distribution (shape/scale parameterization) with matching mean and variance. This yields the following distributions for the downtime duration of the $i$-th (overall) outage that occurred at time $t_{CO,i}$ at service $j$ under the log-normal, resp.\ Gamma, model:
\begin{align}
\theta^{LN}_{CO,i}&:=\theta^{LN}_{CO}(t_{CO,i}) \sim \text{Lognormal}\left(\mu(t_{CO,i}), \sigma^2(t_{CO,i}) \right), \nonumber \\
\mu(t)&=\exp \{-0.105 + 0.119 \lfloor t \rfloor\}, \quad 
\sigma(t)=\exp\{0.482 + 0.018 \lfloor t \rfloor\}, \nonumber \\
& \nonumber \\ 
\theta^{G}_{CO,i}&:=\theta_{CO}^G(t_{CO,i}) \sim \text{Gamma}\left(\kappa(t_{CO,i}), \delta(t_{CO,i}) \right), \nonumber \\
\kappa(t)&=\Bigl( \exp\bigl\{\sigma^2(t)\bigr\} - 1 \Bigr)^{-1}, \quad 
\delta(t)=\exp\bigl\{\sigma^2(t)\bigr\} \exp\left\{ \mu(t) + \frac{\sigma^2(t)}{2} \right\}.  \label{DurationModel}
\end{align}
For later interpretation, observe that the above specification results in a tendency for Gamma-distributed durations to be longer than their log-normal counterparts outside of extreme observations as a direct consequence of equal first and second moments at lower tail weight.
\par 
We employ a linear model to translate simulated outage durations $\theta^{*}_{CO}$, $* \in \{LN,G\}$ into monetary losses. This is motivated by fixed and variable costs caused by an outage event, the latter growing with increasing duration, and prior applications of the linear cost model, see, e.g., \cite{Naldi2017,Franke2020}. Thus, an outage occurring at time $t_{CO,i}$ causes policyholder $k$ to suffer the monetary loss
\begin{equation}
S^{(k), *}_{CO,i} = c_{fix}^{(k)} + c_{var}^{(k)}\, \theta^{*}_{CO,i} + \epsilon_i^{(k),*},
\end{equation}
where $\epsilon_i^{(k),*} \sim \mathcal{N}\left(0,\sigma_{CO,\epsilon}^2\right)$, $\sigma_{CO,\epsilon} > 0$, i.i.d.\ $\forall i,k,*$, describes the independent regression error. Fixed and variable costs $c_{fix}^{(k)}$, resp.\ $c_{var}^{(k)}$, depend on policyholders' characteristics $s^{(k)}$ and $b^{(k)}$ according to assumptions which result\footnote{$\sigma_{CO,\epsilon}$ was chosen based on personal judgment. Its precise value should not influence the qualitative observations of our simulation study, if it is appropriately small.} in (\ref{TrueCost}). Note that \ref{CO6} and \ref{CO7} are motivated qualitatively by Patterson's simple cost estimate, i.e.\ larger companies are expected to have higher average hourly revenues, see \cite{Patterson2002}, and the Ponemon Institute's 2016 \emph{Cost of data center outages} report, see \cite[Bar Chart 4]{Ponemon2016}. 
\begin{enumerate}[start=5, label={(CO\arabic*)}]
\item The expected loss to a policyholder with baseline covariates $s^{(k)} = 1$ and $b^{(k)} = GOV$ for an incident in year $1$ is $\mathbb{E}\big[S^{(k), *}_{CO,i}\big] \approx 40$. It is caused in \emph{equal parts} by fixed and expected variable costs. This amount places us far below the very high cost estimates reported by \cite{Ponemon2016}, i.e.\ about $\$ 9\,000$ per minute of downtime. As these values might be biased towards more extreme events (higher costs) and the results of our simulation study can be expected to exhibit the same qualitative trends under scaling, this choice should not pose any problems for our work.  
\item An increase in size $s^{(k)}$ by one (two) levels causes a growth of 10\% (20\%) in variable costs, but leaves fixed costs unchanged. \label{CO6}
\item Policyholders belonging to one of the sectors $BR$, $MAN$, $EDU$ ($FI$, $HC$) suffer 10\% (20\%) higher fixed costs per outage. Variable costs are unaffected. \label{CO7}
\end{enumerate}
\begin{align} \label{TrueCost}
&c_{fix}^{(k)} = \exp \left\{2.996 + 0.095\,\mathds{1}_{\bigl\{ b^{(k)} \in \{BR,MAN,EDU\}\bigr\}} + 0.18\,\mathds{1}_{\bigl\{ b^{(k)} \in \{FI,HC\}\bigr\}} \right\}, \nonumber \\
&c_{var}^{(k)} = \exp \left\{0.784 + 0.095\,\mathds{1}_{\{ s^{(k)}= 2\}} + 0.18\mathds{1}_{\{ s^{(k)}= 3\}} \right\}, \quad \sigma_{CO,\epsilon} = 2.5.
\end{align}

\subsection{Trigger areas and payment schemes}\label{TriggerPayment}
Based on the above framework, we model pure parametric insurance contracts on the index $\theta_{CO}$ with different risk weightings $\alpha^{(k)} \approx 0.45,0.5,0.55,0.75$ that result in optimal expectile levels $\gamma^{(k)} = 0.4,0.5,0.6,0.9$. Additionally, we differentiate between \emph{static} contracts, where we (wrongfully!) assume outages to be distributed with first year parameters $\mu(0),\sigma(0)$, resp. $\kappa(0),\delta(0)$, over all years, and \emph{dynamic} contracts, where the model (\ref{DurationModel}) is known.
\par
\begin{figure}[h] 
\center
\includegraphics[width = \textwidth, keepaspectratio]{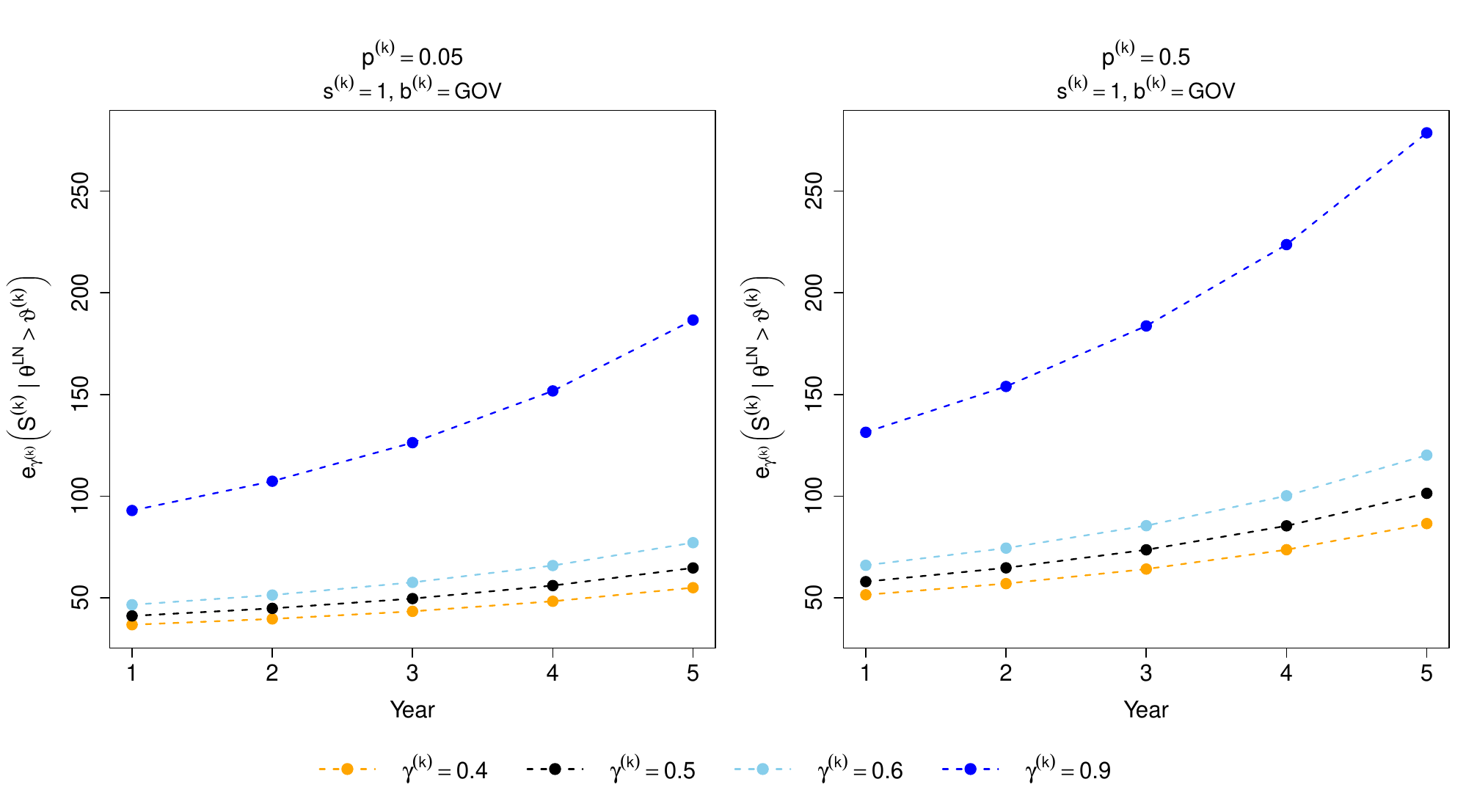}
\caption{Development of the optimal fixed payment (\ref{PayoutSimStudy}) for a baseline policyholder $\left(s^{(k)}=1,b^{(k)} = GOV \right)$ with threshold probability $p^{(k)} = 0.05$ (left), resp.\ $p^{(k)} = 0.5$ (right), for years $1$ to $5$ under log-normally distributed outage durations.} 
\label{OptPaymentTime}
\end{figure}

 
In the pure parametric contracts, a payment for the $i$-th incident is triggered if the observed outage duration $\theta^{*}_{CO,i}$ exceeds $\vartheta^{(k),*}$. Using that outages at different services occur almost surely at distinct times, the trigger area of policyholder $k$ is formally given by $\Theta^{(k),*} = 
 \bigl( (\vartheta^{(k),*},\infty) \times [0,\infty] \times [0,\infty] \bigr) \cup \bigl( [0,\infty] \times (\vartheta^{(k),*},\infty) \times [0,\infty] \bigr) \cup \bigl( [0,\infty] \times [0,\infty] \times (\vartheta^{(k),*},\infty) \bigr)$, where the downtime of a service not suffering an outage in the $i$-th overall incident is treated as zero. $\vartheta^{(k),*}$ is chosen as the $p^{(k)}$-quantile of the underlying outage duration. By linearity of expectiles, see Proposition~\ref{Prop1}, policyholder $k$ receives the optimal fixed compensation
\begin{equation}\label{PayoutSimStudy} 
y^{(k)} = c_{fix}^{(k)} + c_{var}^{(k)}\,  e_{\gamma^{(k)}}\left( \theta^{*}_{CO}\middle| \theta^{*}_{CO} > \vartheta^{(k),*}\right), \,* \in \{LN,G\},
\end{equation}
which is calculated numerically by solving (\ref{ExpectileEq}) using the identities
\begin{align*}
G^{(k),*}_t(x) &:= \int_{-\infty}^x u \, f^{(k),*}_t (u) \, du \\
&= \begin{cases}
0 &, x \le \vartheta^{(k),*}\\
\frac{\exp \left\{\mu(t) + \frac{\sigma^2(t)}{2} \right\}}{1 - p^{(k)}} \Bigl(\Psi_{\mu(t),\sigma(t)}(x) - \Psi_{\mu(t),\sigma(t)}\left(\vartheta^{(k),LN}\right) \Bigr)&,  x > \vartheta^{(k),LN}, * = LN,\\
\frac{\kappa(t) \delta(t)}{1 - p^{(k)} }\left( F^G_{\kappa(t)+1,\delta(t)}(x) - F^G_{\kappa(t)+1,\delta(t)}(\vartheta^{(k),G})\right)&, x > \vartheta^{(k),G},  * = G,
\end{cases} \\
\Psi_{\mu,\sigma}(x) &:= \Phi \left( \frac{\log\left(x\right)- \mu- \sigma^2}{\sigma}\right) \\
E^{(k),*}_t &:= \int_{-\infty}^\infty u \, f^{(k),*}_t (u) \, du \\
&= \begin{cases}
\frac{\exp \left\{\mu(t) + \frac{\sigma^2(t)}{2} \right\}}{1 - p^{(k)}} \Phi \left( \frac{\mu(t) + \sigma^2(t) - \log\left(\vartheta^{(k),LN}\right)}{\sigma(t)}\right) &, \, * = LN,\\
\frac{\kappa(t) \delta(t) }{1 - p^{(k)} } \left(1- F^G_{\kappa(t)+1,\delta(t)}(\vartheta^{(k),G})\right)&, \, * = G,
\end{cases}
\end{align*}
see Appendix~\ref{DerivationExpectiles}, where $f^{(k),*}_t$, $* \in \{LN,G\}$, $k \in\{1,\ldots,K\}$ denotes the conditional density of \mbox{$\theta^{*}_{CO}\big| \theta^{*}_{CO} > \vartheta^{(k),*}$} for an incident at time $t_i \in [t,t+1)$, $\Phi(\cdot)$ is the cdf of $\mathcal{N}(0,1)$, and $F^G_{\kappa(t)+1,\delta(t)}(\cdot)$ is the cdf of a $\text{Gamma}(\kappa(t)+1,\delta(t))$-distribution. The development of the resulting optimal payments over years $1$ to $5$ for a baseline policyholder with $5\%$-, resp.\ $50\%$,-quantile threshold are visualized in Figure~\ref{OptPaymentTime}.

\subsection{Simulation results}\label{SimResults}
To highlight interesting qualitative effects, we consider deviations between actual losses and payouts on a (sub-)portfolio level instead of investigating individual policyholders. In this section, the term ``basis risk'' therefore refers to the sum of differences between payments and experienced losses, i.e.
\begin{equation}\label{BRsim}
g^{(k)}\left(\theta_{CO,i}^*,t_i \right) - S^{(k),*}_{i}, \; * \in \{LN,G\},
\end{equation}
over all policyholders $k$ in the respective (sub-)portfolio. This convention is motivated by matching the sign of (\ref{BRsim}) to the terms \emph{positive} and \emph{negative} basis risk as introduced in Section~\ref{sec3}. Unless indicated otherwise, all values and figures correspond to first-year observations and symmetric risk weighting $\alpha^{(k)} = 0.5$.
\par
\begin{figure}[t] 
\center
\includegraphics[width = \textwidth, keepaspectratio]{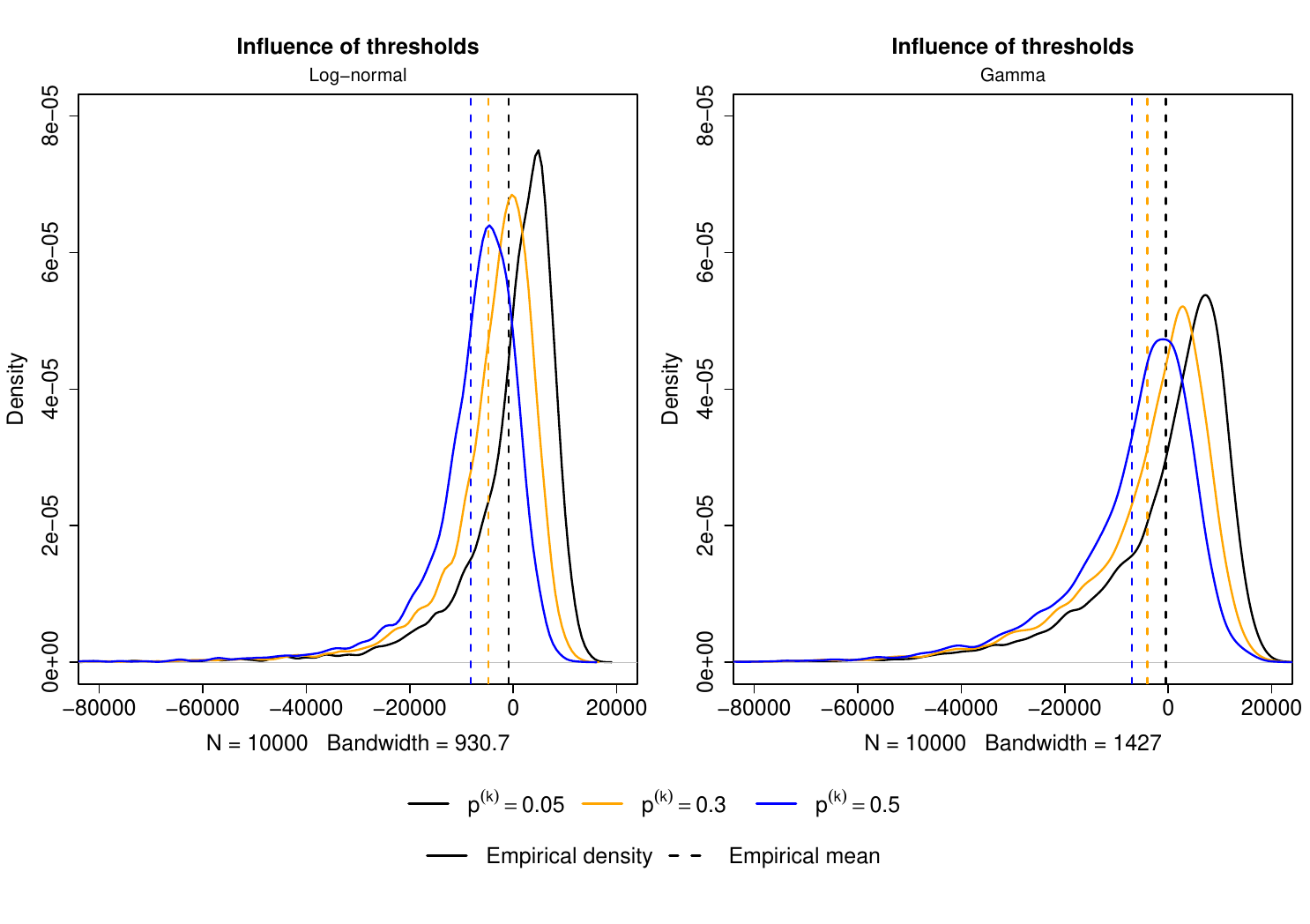}
\caption{Empirical density (solid lines) and mean (dashed lines) of basis risk for threshold sub-portfolios with $p^{(k)} \in \{0.05,0.3,0.5\}$ under log-normally (left) and Gamma-distributed (right) outage durations.} 
\label{THRESHOLD_Basisrisk}
\end{figure}

 
We start by observing the influence of policyholders' threshold quantile levels on basis risk. As expected, an increase in $p^{(k)}$, and thus in $\vartheta^{(k)}$, results in a distributional shift towards negative basis risk, see Figure~\ref{THRESHOLD_Basisrisk}, as a larger proportion of losses goes unpaid. Furthermore, the difference in payments and losses becomes more variable, which is the result of both an increasing number of incidents without compensation and a higher conditional variability of losses.

\begin{figure}[t] 
\center
\includegraphics[width = \textwidth, keepaspectratio]{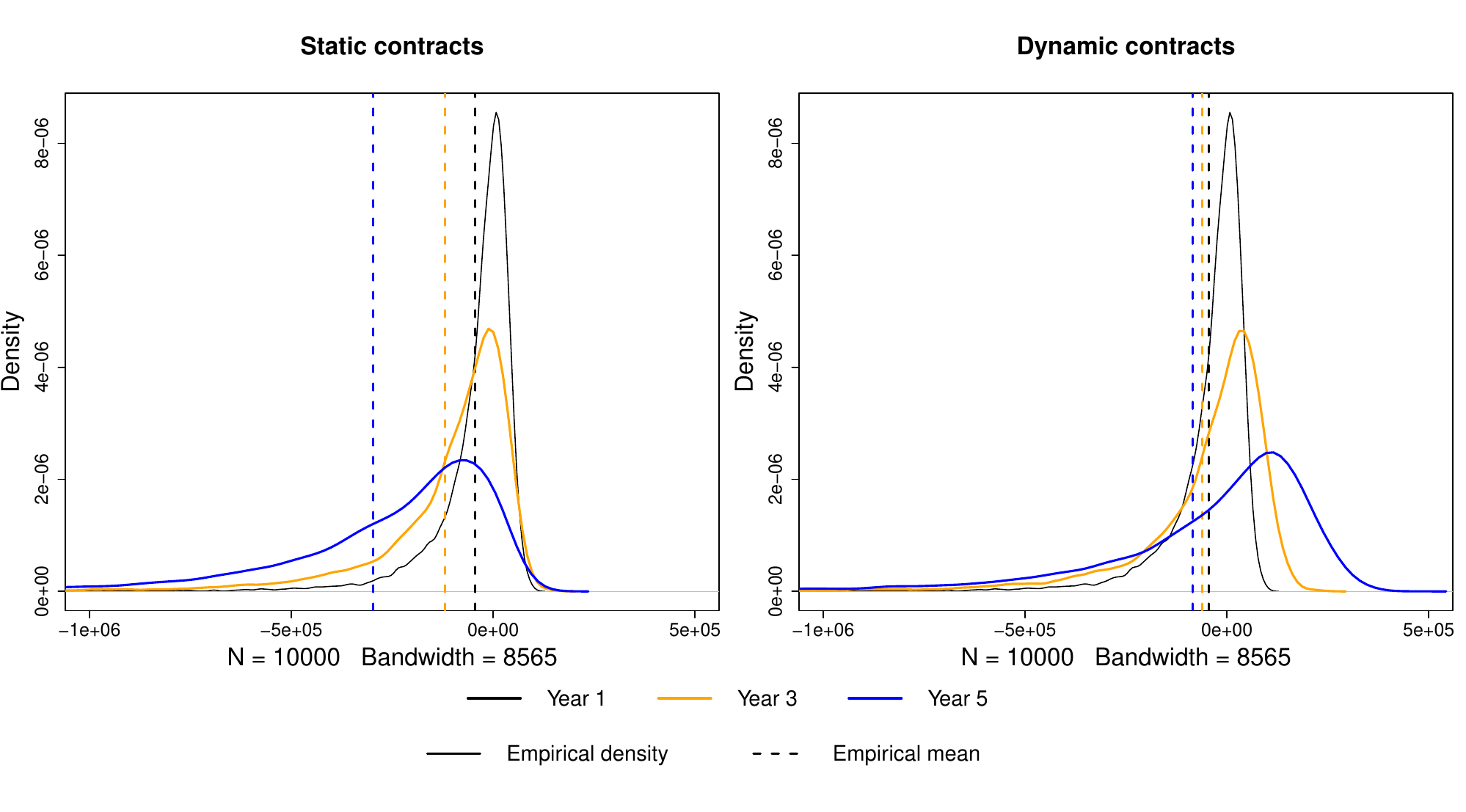}
\caption{Empirical density and mean of basis risk for years $1$, $3$, and $5$ on portfolio level under static (left) and dynamic (right) thresholds and payments. Note that static and dynamic values coincide by definition in year $1$.} 
\label{TIME_Basisrisk}
\end{figure}

Additionally, we can visualize the detrimental effect of ignoring time-dependence through the use of static thresholds and payments in pure parametric insurance. The corresponding basis risk densities in Figure~\ref{TIME_Basisrisk} (left) clearly show a strong shift towards negative basis risk, i.e.\ policyholders face larger residual risk. The use of dynamic thresholds and time-appropriate payments (right) can stabilize average basis risk but cannot reflect the increased variability of losses in a meaningful way, thus still resulting in a worse situation relative to the first year.

\begin{figure}[t] 
\center
\includegraphics[width = \textwidth, keepaspectratio]{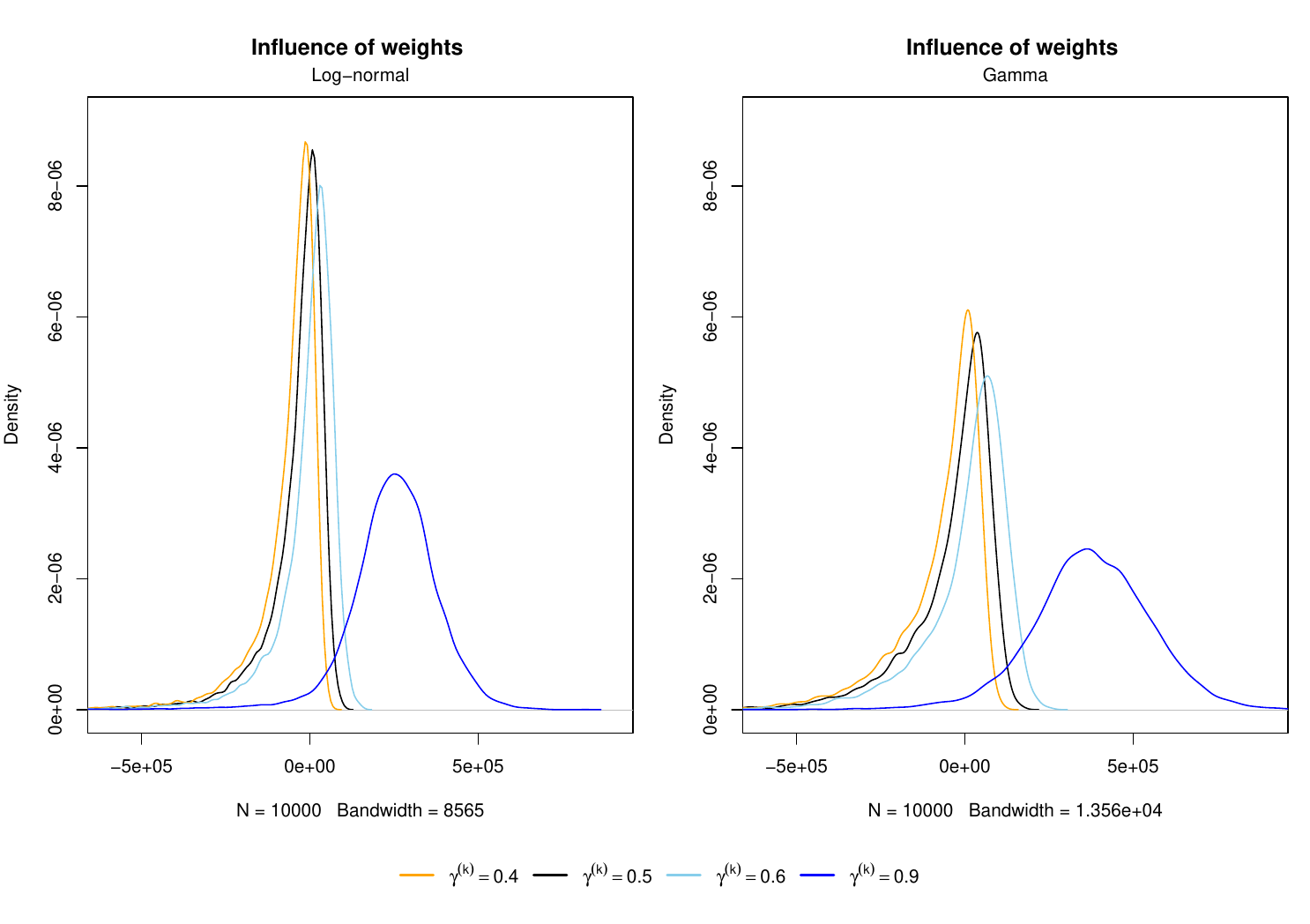}
\caption{Basis risk densities on portfolio level for pure parametric insurance with expectile levels $\gamma^{(k)} =0.4,0.5,0.6,0.9$, corresponds to basis risk weights $\alpha^{(k)} \approx. 0.45,0.5,0.55,0.75$, for log-normally (left) and Gamma-distributed (right) outage durations.} 
\label{WEIGHTS_Basisrisk}
\end{figure}

Next, we can compare basis risk under pure parametric insurance for different risk weightings $\alpha^{(k)}$, resp.\ corresponding expectile levels $\gamma^{(k)}$, used for the calculation of optimal payment. The results in Figure~\ref{WEIGHTS_Basisrisk} can be explained by monotonicity of expectiles w.r.t.\ their level: As $\gamma^{(k)}$ increases, so does the fixed payout $c_{fix}^{(k)} + c_{var}^{(k)} \, e_{\gamma^{(k)}}\left( \theta^{*}_{CO}\middle| \theta^{*}_{CO} > \vartheta^{(k),*}\right)$ for any given threshold $\vartheta^{(k),*}$. Thus, both mean and variance of cumulative payments increase, which results in the shift towards positive basis risk and the widening density. This also explains the differences between the results under $\theta_{CO}^{LN}$ and $\theta_{CO}^{G}$, as the Gamma-distributed durations cause higher payments than their log-normal counterparts if $\gamma^{(k)}$ is large. 
\pagebreak

\begin{figure}[h] 
\center
\includegraphics[width = \textwidth, keepaspectratio]{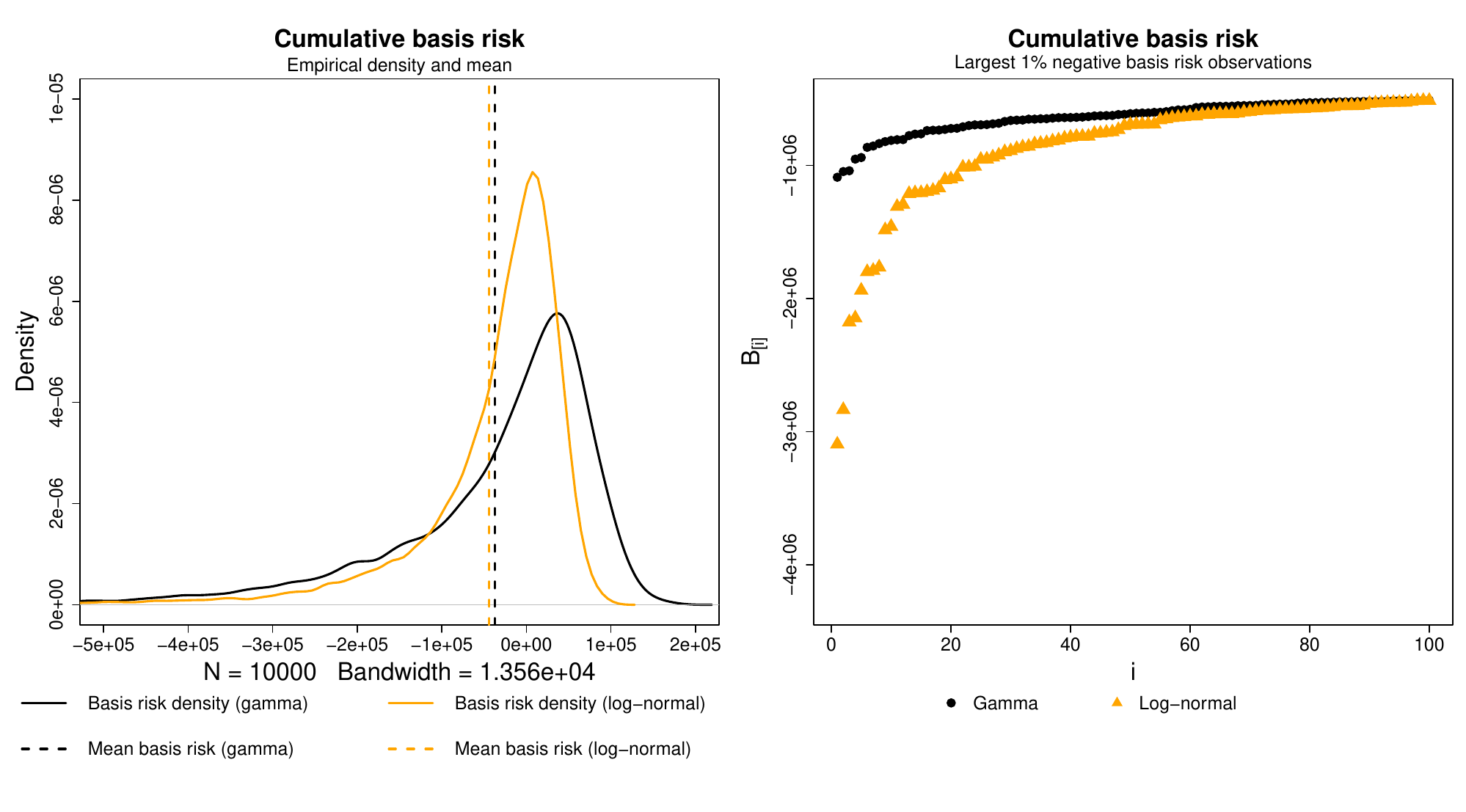}
\caption{Basis risk density and mean (left), resp.\ top $1\%$ largest negative basis risk realizations (right), on portfolio level under log-normally and Gamma-distributed outage durations.} 
\label{TAIL_Basisrisk}
\end{figure}

Finally, we compare the basis risk under log-normally and Gamma-distributed outage durations in Figure~\ref{TAIL_Basisrisk} to assess the impact of tail weight. At first glance, while mean basis risk is approximately equal, Gamma-distributed durations seem to be less preferable with regards to basis risk. We explain this behavior by $\theta_{CO}^G$ being more likely to exhibit high values, see Section~\ref{Severity}. If we focus on the negative basis risk tail, however, which is arguably the most important factor for ensuring policyholder retention and thus long-term viability of parametric insurance, we observe the opposite effect. As can easily be seen from the largest $1\%$ negative basis risk realizations, the increased tail-weight of log-normally distributed outage durations results in far more extreme coverage gaps, leaving policyholders severely under-protected. Thus, we confirm \cite{Lopez2022} in noting the difficulties in protecting policyholders against heavy-tailed damages through parametric insurance.

\section{Conclusion and outlook}\label{sec5}
Parametric insurance is a risk-transfer solution that indemnifies policyholders via a payment that is determined by an index. This payment is considered as proxy for the true loss experienced. A major hindrance for the implementation and acceptance of parametric coverage is the presence of basis risk, which arises when policyholders' compensation deviates from the actual loss. This risk is typically measured in terms of either the difference between the two quantities or the correlation between losses and the chosen index. For explicit minimization of basis risk through appropriate contract design, existing research has largely focused on specific risk scenarios. In contrast, to the best of our knowledge, no analytic expression for the optimal design of basis risk-minimizing payment schemes under a given generic index has been provided so far and this is were our contribution lies.
\par 
To reduce the basis risk associated with parametric insurance contracts, we introduced a quantification framework by asymmetrically weighting the positive and negative squared differences between true losses and payouts. This not only connects our framework to existing approaches (MSE, relative residual variance, basic penalty functions) but also accounts for important characteristics of basis risk. Most striking, however, it transfers the problem to the domain of (conditional) expectiles. In this setting, we are able to derive the optimal fixed payment to a policyholder under pure parametric insurance as a conditional expectile of their true loss given payment being triggered. In the more general case of variable payments under parametric index insurance, the optimal compensation is again a conditional expectile of the true loss, now taking into account the concrete index realization and incident time. The expectile level is equal in both cases and depends on the importance of negative basis risk relative to its positive counterpart. Payment schemes of aggregate loss index insurance contracts, such as the commonly implemented area-yield crop insurance, can be treated as a special case.
\par 
Our optimal payment schemes are shown to have intuitive properties. As our results were derived in a general setting, they can be combined with existing considerations on ideal index and trigger design. This supports a holistic approach to minimizing basis risk. Additionally, using known characteristics of (conditional) expectiles, we analyzed the influence of risk perception/measurement and stochastic orderings of the conditional true losses on the amount of compensation under the optimal payment scheme. Further, expectile regression yields an efficient solution for practical implementation and provides a link to commonly used compensation schemes like linear or step functions. 
\par 
A simulation study on the transfer of risk arising from cloud outages visualizes a potential application in the context of cyber insurance and supports both our theoretical observations as well as existing results on the dangers of basis risk under heavy-tailed loss distributions.
\par 
While these results help minimizing basis risk in parametric insurance, the search for a ``perfect'' product design is by no means at an end. Specifically, let us mention three topics that we believe provide room for future research on expectile-based payment schemes. First, we currently assume the relative importance $\alpha^{(k)}$ of negative basis risk to be known, which possibly is too optimistic in practical applications. Thus, a way to reduce the uncertainty inherent to our approach would be an important step towards increasing the robustness of the proposed payment scheme design, ideas from behavior insurance might apply here. Second, our current measurement of basis risk only accounts for the absolute difference of losses and payments. However, as motivated in the context of basic penalty functions, these deviations should also be considered relative to the overall payout amount, see \cite{Claramunt2022}. Third, prior results on parametric contract design, using different basis risk quantification methods and considerations on index design, see, e.g., \cite{Stigler2023}, could be revisited under our framework, possibly aligning their results \pagebreak 
with our minimization approach. Naturally, we expect frameworks based on the MSE or the closely related relative residual variance metric, see Table~\ref{BasisRiskMeasures}, to be particularly suited as starting points in the search for a holistic design of basis risk-optimal parametric insurance.

\begin{appendix}

\section{Proofs and derivations}

\subsection{Proof of Proposition \ref{OptBRpure}}\label{ProofPureParametric}
For ease of notation, the superscript $(k)$ is omitted in the following. Additionally, let $\Theta \subset \theta(\Omega)$ such that w.l.o.g.\ $\mathbb{P}(\theta \in \Theta) \in (0,1)$\footnote{If $\mathbb{P}(\theta \in \Theta) = 1$, $Y = y$ a.s.\ would allow us to move directly to solving (\ref{MinStepPPproof}) without making use of the tower property. Since $\theta \in \Theta$ a.s., $\mathbb{P}(\circ|\theta \in \Theta) = \mathbb{P}$ and we find the \emph{unconditional} $\gamma^{(k)}$-expectile to be the unique minimizer, analogously to the presented proof.}. With the tower property and (\ref{PSpure}):
\begin{align}\label{proofBRpureeq1}
\mathbb{E}[B] =& \mathbb{E}\left[\alpha^2 \Bigl(\bigl(S - Y \bigr)^+\Bigr)^2 + (1-\alpha)^2  \Bigl(\bigl(S - Y \bigr)^-\Bigr)^2\right] \nonumber \\
=& \mathbb{P}(\theta \in \Theta)\mathbb{E}\left[\alpha^2 \Bigl(\bigl(S - Y \bigr)^+\Bigr)^2 + (1-\alpha)^2  \Bigl(\bigl(S - Y \bigr)^-\Bigr)^2 \middle| \theta \in \Theta \right]\nonumber\\
& + \mathbb{P}(\theta \notin \Theta)\mathbb{E}\left[\alpha^2 \Bigl(\bigl(S - Y \bigr)^+\Bigr)^2 + (1-\alpha)^2  \Bigl(\bigl(S - Y \bigr)^-\Bigr)^2 \middle| \theta \notin \Theta \right]\nonumber\\
=&\mathbb{P}(\theta \in \Theta) \left(\alpha^2 \mathbb{E}\left[ \Bigl(\bigl(S - y \bigr)^+\Bigr)^2 \middle| \theta \in \Theta \right] + (1-\alpha)^2 \mathbb{E} \left[ \Bigl(\bigl(S - y \bigr)^-\Bigr)^2 \middle| \theta \in \Theta \right] \right)\nonumber\\
& + \mathbb{P}(\theta \notin \Theta) \alpha^2 \mathbb{E}\left[S^2 \middle| \theta \notin \Theta \right].
\end{align}
Thus, our minimization simplifies to solving
\begin{equation}\label{MinStepPPproof}
\underset{y > 0}{\mathrm{argmin}} \left(\alpha^2 \mathbb{E}\left[ \Bigl(\bigl(S - y \bigr)^+\Bigr)^2 \middle| \theta \in \Theta \right] + (1-\alpha)^2 \mathbb{E} \left[ \Bigl(\bigl(S - y \bigr)^-\Bigr)^2 \middle| \theta \in \Theta \right] \right).
\end{equation}
Setting $\gamma = \frac{\alpha^2}{(1 - \alpha)^2 + \alpha^2}$, i.e.\ $\frac{1 - \gamma}{\gamma} = \frac{(1-\alpha)^2}{\alpha^2}$, one can easily check that $\gamma \in (0,1)$ is well-defined for all  $\alpha \in (0,1)$. Further, $(1-2\alpha + 2\alpha^2) \ge (1-2 \cdot 0.5 + 2\cdot 0.5^2) = 0.5>0$ for any $\alpha \in (0,1)$ follows through trivial calculus. Thus, (\ref{MinStepPPproof}) is equivalent to
\begin{equation*}
\underset{y > 0}{\mathrm{argmin}} \left(\gamma \mathbb{E}\left[ \Bigl(\bigl(S - y \bigr)^+\Bigr)^2 \middle| \theta \in \Theta \right] + (1-\gamma) \mathbb{E} \left[ \Bigl(\bigl(S - y \bigr)^-\Bigr)^2 \middle| \theta \in \Theta \right] \right),
\end{equation*}
which is by Definition~\ref{DefCExp} uniquely solved by $y = e_\gamma \left(S| \theta \in \Theta\right)$. This proves (\ref{proof1}). As $\alpha = 0.5$ implies $\gamma = 0.5$ and $e_{0.5} \left(S| \theta \in \Theta\right) = \mathbb{E}\left[ S| \theta \in \Theta \right]$ is an obvious consequence of Definition~\ref{DefCExp}, inserting $y = e_{0.5} \left(S| \theta \in \Theta\right)$ in (\ref{proofBRpureeq1}) verifies (\ref{proof2}).

\subsection{Proof of Proposition \ref{OptBRindex}}\label{ProofParametricIndex}
Again, all superscripts are omitted. Let $\alpha \in (0,1)$. A tower property argument and rearranging terms, analogous to the proof of Proposition~\ref{OptBRpure}, show that minimizing (\ref{ExpBRindex}) over all permissible $Y$ is equivalent to solving
\begin{align*}
 \tilde{g}(\theta,\tau)&=\underset{Y\in \mathcal{L}^2 (\Omega, \sigma(\theta,\tau), \mathbb{P}), Y \ge 0 \, \mathrm{a.s.}}{\mathrm{argmin}} \, \mathbb{E}\left[ \gamma \left(  \bigl(S- Y \bigr)^+\right)^2 +(1 - \gamma) \left( \bigl(S - Y \bigr)^-  \right)^2 \middle| \theta \in \Theta \right] \nonumber \\
&=\underset{Y \in \mathcal{L}^2 (\Omega, \sigma(\theta,\tau), \mathbb{P}(\circ| \theta \in \Theta)), Y\ge 0 \, \mathrm{a.s.}}{\mathrm{argmin}} \, \mathbb{E}\left[ \gamma \left(  \bigl(S- Y \bigr)^+\right)^2 +(1 - \gamma) \left( \bigl(S - Y \bigr)^-  \right)^2 \right].
\end{align*}
Thus, we can directly conclude 
\begin{align*}
\tilde{g}(\theta,\tau) &= e_{\gamma}\left(S|\theta,\tau;\theta \in \Theta \right) \\
&:=\underset{Y \in \mathcal{L}^2 (\Omega, \sigma(\theta,\tau), \mathbb{P}(\circ| \theta \in \Theta))}{\mathrm{argmin}} \, \mathbb{E}\left[ \gamma \left(  \bigl(S- Y \bigr)^+\right)^2 +(1 - \gamma) \left( \bigl(S - Y \bigr)^-  \right)^2 \right]
\end{align*}
by Definition~\ref{DefCExp}. Note that Lemma~\ref{PropCExp} $\mathbb{P}(\circ| \theta \in \Theta)$-a.s.\ guarantees $e_{\gamma}\left(S|\theta,\tau; \theta \in \Theta \right) \ge 0$. To conclude (\ref{proof1Index}), it is sufficient by Lemma~\ref{PropCExp} to show equality of the regular conditional distribution of $S$ given $(\theta,\tau)$ under $\mathbb{P}$ and $\tilde{\mathbb{P}} := \mathbb{P}(\circ|\theta \in \Theta)$ for any $(\theta_i,t_i) \in \Theta \times [0,T]$. While intuitive, we rigorously derive this equality: Let $A \in \mathcal{B}(\mathbb{R}_{\ge 0})$ and $(\theta_i,t_i) \in \Theta \times [0,T]$. By definition of the regular conditional distribution, see, e.g., \cite[Definition 8.28]{Klenke2020}, it suffices to prove 
\begin{equation} \label{CondDistrProof}
\mathbb{P}(S \in A| \theta = \theta_i, \tau = t_i )= \tilde{\mathbb{P}}(S \in A| \theta = \theta_i, \tau = t_i).
\end{equation}
For $X_A:= \mathds{1}_{\{S \in A\} }$, $Z_A:= \mathbb{E}_{\mathbb{P}} [X_A|\theta,\tau]$, and $\tilde{Z}_A:= \mathbb{E}_{\tilde{\mathbb{P}}}[X_A|\theta,\tau]$, it holds for any $B \in \sigma(\theta,\tau)$
\begin{align*}
\mathbb{E}_{\tilde{\mathbb{P}}}\left[X_A\,\mathds{1}_{B}\right] &= \frac{\mathbb{E}_\mathbb{P} \left[X_A\,\mathds{1}_{B}\, \mathds{1}_{\{\theta \in \Theta\} }\right]}{\mathbb{P}(\theta \in \Theta)} = \frac{\mathbb{E}_\mathbb{P} \left[X_A\,\mathds{1}_{B \cap \{\theta \in \Theta\} }\right]}{\mathbb{P}(\theta \in \Theta)} \\
&\overset{(*)}{=} \frac{\mathbb{E}_\mathbb{P} \left[Z_A\,\mathds{1}_{B \cap \{\theta \in \Theta\} }\right]}{\mathbb{P}(\theta \in \Theta)} = \frac{\mathbb{E}_\mathbb{P} \left[Z_A\,\mathds{1}_{B}\, \mathds{1}_{\{\theta \in \Theta\} }\right]}{\mathbb{P}(\theta \in \Theta)} =\mathbb{E}_{\tilde{\mathbb{P}}}\left[Z_A\,\mathds{1}_{B}\right],
\end{align*}
where $(*)$ follows by definition of $Z_A$, since $\{\theta \in \Theta\} = \bigl\{ (\theta,\tau) \in \Theta \times \mathbb{R} \bigr\} \subset \sigma(\theta,\tau)$. Now, by uniqueness of the conditional expectation, $Z_A = \tilde{Z}_A$ $\tilde{\mathbb{P}}$-almost everywhere. Thus, using the factorization lemma, there exists $\phi: \Theta \times [0,T] \to \mathbb{R}$ s.t. $\phi(\theta,\tau) = Z_A = \tilde{Z}_A$ $\tilde{\mathbb{P}}$-a.s. and 
\begin{equation*}
\mathbb{P}(S \in A| \theta = \theta_i, \tau = t_i )= \phi(\theta_i,t_i) = \tilde{\mathbb{P}}(S \in A| \theta = \theta_i, \tau = t_i),
\end{equation*}
see, e.g., \cite[Definition 8.24]{Klenke2020}. Since $A$ was chosen arbitrarily, this proves (\ref{CondDistrProof}).
\par 
With regards to (\ref{proof2Index}), we note that $\alpha = 0.5$ implies $\gamma = 0.5$ and $e_{0.5}\left(S|\theta,\tau\right) = \mathbb{E}\left[S \middle| \theta,\tau \right]$ a.s.\ is a consequence of Definition~\ref{DefCExp}. The rest of the proof follows directly from
\begin{align*} 
&\mathbb{E}\left[ \alpha^2 \left(  \bigl(S- g_{opt}(\theta,\tau) \bigr)^+\right)^2 +\bigl(1 - \alpha\bigr)^2 \left( \bigl(S - g_{opt}(\theta,\tau) \bigr)^-  \right)^2 \right] \nonumber \\
&= \frac{1}{4}\left( \mathbb{P}(\theta \in \Theta)\mathbb{E}\Bigl[\mathbb{E}[ \bigl( S - \mathbb{E}\left[S \middle| \theta,\tau \right] \bigr)^2| \theta,\tau ] \Big|\, \theta \in \Theta \Bigr] + \mathbb{P}(\theta \notin \Theta)\mathbb{E}\left[ S^2 \middle| \, \theta \notin \Theta \right] \right).
\end{align*}

\subsection{Proof of Corollary \ref{LossWOIncident}}\label{ProofLossWOIncident}
Using the tower property and omitting all superscripts, we write
\begin{equation*}
\mathbb{E}\left[B\right] = p\,\alpha^2 \mathbb{E}\left[  S^2 \middle| Z = 0 \right] + \left(1-p \right)\mathbb{E}\left[ B \middle| Z = 1 \right].
\end{equation*}
Thus, minimizing expected basis risk in the setting of Proposition~\ref{OptBRpure}, resp.\ \ref{OptBRindex}, is equivalent to finding 
\begin{equation*}
\underset{Y}{\mathrm{argmin}} \, \mathbb{E}\left[ \alpha^2 \bigl( (S - Y)^+ \bigr)^2 + \left(1-\alpha \right)^2  \bigl((S - Y )^-\bigr)^2 \middle| Z = 1 \right],
\end{equation*}
for appropriate $Y$ and the respective proofs follow analogously to \ref{ProofPureParametric} and \ref{ProofParametricIndex} under $\mathbb{P}(\circ|Z = 1)$ instead of $\mathbb{P}$.

\subsection{Example \ref{AreaYieldExample} - True loss distribution}\label{AppYieldLossDistr} 
Let $k \in \{1,\ldots,K\}$ and 
\begin{equation*}
\left(C^{(1)}, \ldots, C^{(K)}\right)' \sim \mathcal{N} \left(\nu, \Sigma \right),
\end{equation*} 
with 
\begin{align*}
\nu &= (\mu \rule[0.5ex]{1.5em}{0.4pt} \mu)' \in \mathbb{R}^{K\times 1},\quad \Sigma = \Biggl( \sigma^2 \mathrm{Corr}\Bigl(C^{(i)},C^{(j)} \Bigr) \Biggr)_{(i,j) \in \{1,\ldots,K\}^2} \in \mathbb{R}^{K\times K}.
\end{align*}
By familiar properties of the multivariate normal distribution, the vector
\begin{equation*}
\begin{pmatrix}
C^{(k)} \\
\theta
\end{pmatrix} =\underbrace{ \begin{pmatrix}
e_{[k]}'\\
\boldsymbol{1}'
\end{pmatrix}}_{A:=}
\begin{pmatrix}
C^{(1)}\\
\vdots \\
C^{(K)}
\end{pmatrix}
\end{equation*}
is again (bivariate) normally distributed with mean vector
\begin{equation*}
\tilde{\nu} = \begin{pmatrix}
\tilde{\nu}_1\\
\tilde{\nu}_2
\end{pmatrix} = A\nu = \begin{pmatrix}
\mu\\
K\mu
\end{pmatrix},
\end{equation*}
and covariance matrix
\begin{equation*}
\tilde{\Sigma} = \begin{pmatrix}
\tilde{\sigma}_{11} & \tilde{\sigma}_{12} \\
\tilde{\sigma}_{21} & \tilde{\sigma}_{22}
\end{pmatrix} = A\Sigma A' = \begin{pmatrix}
\sigma^2 &\boldsymbol{1}'\Sigma e_{[k]} \\
\boldsymbol{1}'\Sigma e_{[k]} & \boldsymbol{1}'\Sigma \boldsymbol{1}
\end{pmatrix}.
\end{equation*}
Using results for conditioning multivariate normal distributions, we conclude
\begin{align}
&C^{(k)} | \theta = \theta_i \sim \mathcal{N}\left(\mu^*, (\sigma^*)^2\right)  \label{YIELDcdfyield} \\
&\mu^* = \tilde{\nu}_1 + \frac{\tilde{\sigma}_{12}}{\tilde{\sigma}_{22}}(\theta_i - \tilde{\nu}_2) = \mu + \frac{\boldsymbol{1}'\Sigma e_{[k]} }{\boldsymbol{1}'\Sigma \boldsymbol{1}} \left( \theta_i - K\mu \right) \in \mathbb{R}, \nonumber \\
&\sigma^*= \sqrt{ \tilde{\sigma}_{11} - \frac{(\tilde{\sigma}_{12})^2}{\tilde{\sigma}_{22}}}
= \sqrt{ \sigma^2 - \frac{\left( \boldsymbol{1}'\Sigma e_{[k]} \right)^2}{ \boldsymbol{1}'\Sigma\boldsymbol{1}} } \in \mathbb{R}_{>0}, \nonumber
\end{align}
and the result follows from the observations
\begin{align*}
&S^{(k)} \ge 0 \; \mathbb{P}\mathrm{-a.s.}, \, S^{(k)} \le s_{max} \; \mathbb{P}\mathrm{-a.s.}, \\
&S^{(k)} \le s \Leftrightarrow c_{crit} - C^{(k)} \le s \Leftrightarrow c_{crit} - s \le C^{(k)}, \; \forall s \in \left[0, s_{max} \right).
\end{align*}

\subsection{Example \ref{AreaYieldExample} - Derivation of conditional expectiles}\label{AppYieldLossExpectile}
We again omit all superscripts. Recalling (\ref{ExpectileEq}) and our derivations in Appendix~\ref{AppYieldLossDistr}, it is sufficient to find expressions for
\begin{align*}
G(s) &= \int_{-\infty}^s t \, dF_{S|\theta=\theta_i}(t), \quad E  =  \int_{-\infty}^\infty t \, dF_{S|\theta=\theta_i}(t) = \mathbb{E}\left[S\middle| \theta = \theta_i \right],
\end{align*}
for any $\theta_i < \vartheta$. Since $S$ is a deterministic function of $C$, see (\ref{TrueLossAggII}), and $C| \theta = \theta_i$ is normally distributed, see (\ref{YIELDcdfyield}), we can easily see 
\begin{align*}
E  = & \, 0 \cdot \mathbb{P}\left( C \ge c_{crit} \middle| \theta = \theta_i \right) + s_{max} \, \mathbb{P}\left( C \le c_{min} \middle| \theta = \theta_i \right) \\
& + \mathbb{E}[c_{crit} - C | c_{min} < C < c_{crit}, \, \theta = \theta_i]\, \mathbb{P}\left( c_{min} < C < c_{crit}\middle| \theta = \theta_i \right) \\
=& \, s_{max} \, \Phi_{\mu^*,\sigma^*}(c_{min}) + \int_{c_{min}}^{c_{crit}} \left(c_{crit} - t\right) \frac{1}{\sigma^* \sqrt{2\pi}} \exp \left\{ -\frac{(t - \mu^*)^2}{2(\sigma^*)^2}\right\} \, dt \\
\overset{(*)}{=}& s_{max}\, \Phi_{\mu^*,\sigma^*}\left(c_{min}\right) + \int_{\frac{c_{min} - \mu^*}{\sigma^*}}^{\frac{c_{crit} - \mu^*}{\sigma^*}}  \frac{c_{crit}-\mu^* - \sigma^* z}{\sqrt{2\pi}} \exp \left\{ -\frac{z^2}{2}\right\} \, dz \\
=& \, s_{max}\, \Phi_{\mu^*,\sigma^*}\left(c_{min}\right) + \left(c_{crit} - \mu^* \right)\bigl(\Phi_{\mu^*,\sigma^*}\left(c_{crit}\right) - \Phi_{\mu^*,\sigma^*}\left(c_{min}\right) \bigr) \\
&+ \frac{\sigma^*}{\sqrt{2\pi}} \left( \exp \left\{- \frac{(c_{crit} - \mu^*)^2}{2(\sigma^*)^2} \right\} - \exp \left\{- \frac{(c_{min} - \mu^*)^2}{2(\sigma^*)^2}  \right\} \right) \\
=& \left(c_{crit} - \mu^* \right)\Phi_{\mu^*,\sigma^*}\left(c_{crit}\right) + \left(\mu^*- c_{min}\right) \Phi_{\mu^*,\sigma^*}\left(c_{min}\right) \\
&+  \frac{\sigma^*}{\sqrt{2\pi}} \left( \exp \left\{- \frac{(c_{crit} - \mu^*)^2}{2(\sigma^*)^2} \right\} - \exp \left\{- \frac{(c_{min} - \mu^*)^2}{2(\sigma^*)^2}  \right\} \right),
\end{align*}
where $(*)$ follows from substituting $z = (t - \mu^*)/\sigma^*$.
\par 
Similarly, we obtain for any $s \in \left( 0, s_{max} \right)$:
\begin{align*}
G(s) = & \, 0 \cdot \mathbb{P}\left( C \ge c_{crit} \middle| \theta = \theta_i \right) +  \int_{c_{crit} - s}^{c_{crit}} \left(c_{crit} - t\right) \frac{1}{\sigma^* \sqrt{2\pi}} \exp \left\{ -\frac{(t - \mu^*)^2}{2(\sigma^*)^2}\right\} \, dt \\
=&  \left(c_{crit} - \mu^* \right) \bigl(\Phi_{\mu^*,\sigma^*}\left(c_{crit}\right) - \Phi_{\mu^*,\sigma^*}\left(c_{crit} - s \right) \bigr) \\
&+ \frac{\sigma^*}{\sqrt{2\pi}} \left( \exp \left\{- \frac{(c_{crit} - \mu^*)^2}{2(\sigma^*)^2} \right\} - \exp \left\{- \frac{(c_{crit} - s - \mu^*)^2}{2(\sigma^*)^2}  \right\} \right).
\end{align*}
As further $(S|\theta=\theta_i) \in \left[ 0, s_{max} \right]$ $\mathbb{P}$-a.s., we can directly conclude $G(0) = 0$ as well as $G\left( s_{max}\right) = E$. Note that $\lim_{s \uparrow s_{max} } G(s) \neq E$ since $\mathbb{P}(S =s_{max}|\theta = \theta_i) >0$. Now, we have derived all quantities needed to approximate the expectiles of interest using bisection search.

\subsection{Optimal fixed cloud outage compensation}\label{DerivationExpectiles}
Let $x \in \mathbb{R}$. Again, we omit the subscript $CO$ and the superscript $(k)$. Further, we fix time $t \in [0,T]$ and write
$\mu = \mu(t)$, $\sigma = \sigma(t)$, $\kappa = \kappa(t)$, and $\delta = \delta(t)$ for the corresponding time-dependent parameters defined in (\ref{DurationModel}).
\par 
First, we note that $\theta^{LN} \big| \theta^{LN} > \vartheta^{LN}$ has a truncated log-normal distribution. Thus, its density is given by
\begin{align*}
f^{LN}_t(u) &= \frac{
\frac{1}{u\sigma\sqrt{2\pi}}\exp \left\{-\frac{(\log(u) - \mu)^2}{2\sigma^2}\right\}
}{1 - \mathbb{P}\left( \theta^{LN} \le \vartheta^{LN} \right)} \mathds{1}_{\{u > \vartheta^{LN}\}} \\
&= \frac{
\frac{1}{u\sigma\sqrt{2\pi}}\exp \left\{-\frac{(\log(u) - \mu)^2}{2\sigma^2}\right\}
}{1 - p} \mathds{1}_{\{u > \vartheta^{LN}\}},
\end{align*}
since $\vartheta^{LN}$ is the $p$-quantile of $\theta^{LN}$. Thus, it obviously holds
\begin{equation*}
G^{LN}_t(x) = 0, \; \forall x \le \vartheta^{LN},
\end{equation*}
and
\begin{align*} 
G^{LN}_t(x) &= \frac{1}{1-p} \int_{\vartheta^{LN}}^x u \frac{1}{u\sigma\sqrt{2\pi}}\exp \left\{-\frac{(\log(u) - \mu)^2}{2\sigma^2}\right\} \, du \\
&\overset{(*)}{=} \frac{1}{1-p} \int_{\log \left(\vartheta^{LN}\right)}^{\log(x)} \frac{1}{\sigma\sqrt{2\pi}}\exp \left\{z-\frac{(z - \mu)^2}{2\sigma^2}\right\} \, dz \\
&= \frac{\exp \left\{\mu + \frac{\sigma^2}{2}\right\}}{1-p} \int_{\log \left(\vartheta^{LN}\right)}^{\log(x)} \frac{1}{\sigma\sqrt{2\pi}}\exp \left\{-\frac{(z - \mu - \sigma^2)^2}{2\sigma^2}\right\} \, dz \\
&= \frac{\exp \left\{\mu + \frac{\sigma^2}{2} \right\}}{1-p} \Biggl(\Phi \left( \frac{\log(x) - \mu - \sigma^2}{\sigma}\right) - \Phi \left( \frac{\log\left(\vartheta^{LN}\right) - \mu - \sigma^2}{\sigma}\right)\Biggr), \; \forall x > \vartheta^{LN},
\end{align*}
where $(*)$ follows by substituting $z = \log(u)$. The expression for $E^{LN}_t$ follows directly from $E^{LN}_t= \lim_{x\to \infty} G^{LN}_t(x)$ and $1 - \Phi(x) = \Phi(-x)$. 
\par 
Similarly, $\theta^{G} \big| \theta^{G} > \vartheta^{G}$ has a truncated Gamma distribution, whose density is
\begin{equation*}
f^{G}_t(u) = \frac{
\frac{1}{\Gamma(\kappa) \delta^{\kappa}}u^{\kappa - 1}\exp\left(-\frac{u}{\delta}\right)
}{1 - p} \mathds{1}_{\{u > \vartheta^{G}\}}.
\end{equation*}
This again yields 
\begin{equation*}
G^{G}_t(x) = 0, \; \forall x \le \vartheta^{G},
\end{equation*}
and
\begin{align*} 
G^{LN}_t(x) &= \frac{1}{1-p} \int_{\vartheta^{G}}^x u \frac{1}{\Gamma(\kappa) \delta^{\kappa}}u^{\kappa - 1}\exp\left\{-\frac{u}{\delta}\right\} \, du \\
&= \frac{1}{1-p} \frac{\Gamma(\kappa+1) \delta^{\kappa+1}}{\Gamma(\kappa) \delta^{\kappa}} \int_{\vartheta^{G}}^x \frac{1}{\Gamma(\kappa + 1) \delta^{\kappa + 1}}u^{\kappa}\exp\left\{-\frac{u}{\delta}\right\} \, du \\
&= \frac{\kappa \delta}{1-p} \left( F^G_{\kappa+1,\delta}(x) - F^G_{\kappa+1,\delta}\bigl(\vartheta^G\bigr) \right) , \; \forall x > \vartheta^{G},
\end{align*}
using $\Gamma(\kappa+1) = \kappa \Gamma(\kappa)$. $E^{G}_t= \lim_{x\to \infty} G^{G}_t(x)$ concludes our derivations.
\end{appendix}



\end{document}